\documentclass[twocolumn,aps,prl,longbibliography]{revtex4-2}
        \usepackage[utf8]{inputenc}
        \usepackage[T1]{fontenc}
	\usepackage[usenames,dvipsnames]{xcolor}
	\usepackage{amsmath}
	\usepackage{amsfonts}
	\usepackage{amssymb}
	\usepackage[colorlinks=true,citecolor=blue,linkcolor=blue,urlcolor=blue]{hyperref}
	\usepackage{graphicx}
	\usepackage{bbold}					% for \mathbb{1} as a nice unit matrix symbol
	\usepackage[makeroom]{cancel}		% crossed terms in tables
	\usepackage{multirow}				% merge cells vertically in tables
	\usepackage[normalem]{ulem}        % strike-through text
	\usepackage{array}
	\usepackage{pdfpages}
    \usepackage{float}
    \usepackage{upgreek}

	\makeatletter
	\AtBeginDocument{\let\LS@rot\@undefined}
	\makeatother
	
	\newcolumntype{x}[1]{>{\centering\let\newline\\\arraybackslash\hspace{0pt}}p{#1}}

	\DeclareMathAlphabet{\mathbbold}{U}{bbold}{m}{n}

	\newcounter{subeqn} %

	\makeatletter
	\@addtoreset{subeqn}{equation}
	\makeatother

\definecolor{TB}{rgb}{0,0,0} %uncomment for all black

\def\beq{\begin{equation}}
\def\eeq{\end{equation}}
\def\bald{\begin{aligned}}
\def\eald{\end{aligned}}
\def\bea{\begin{eqnarray}}
\def\eea{\end{eqnarray}}

\def\Eq#1{Eq.~(\ref{#1})}
\def\Fig#1{Fig.~\ref{#1}}

\begin{document}

\title{Skin Effect Induced Anomalous Dynamics from Charge-Fluctuating Initial States}

\author{Sibo Guo$^{1,2}$}
\author{Shuai Yin$^{3,4}$}
\author{Shi-Xin Zhang$^{1}$}
\email{shixinzhang@iphy.ac.cn}
\author{Zi-Xiang Li$^{1,2}$}
\email{zixiangli@iphy.ac.cn}

\affiliation{$^1$Beijing National Laboratory for Condensed Matter Physics $\&$ Institute of Physics, Chinese Academy of Sciences, Beijing 100190, China}
\affiliation{$^2$University of Chinese Academy of Sciences, Beijing 100049, China}
\affiliation{$^3$Guangdong Provincial Key Laboratory of Magnetoelectric Physics and Devices, School of Physics, Sun Yat-Sen University, Guangzhou 510275, China}
\affiliation{$^4$School of Physics, Sun Yat-Sen University, Guangzhou 510275, China}

\date{\today}

\begin{abstract}
Non-equilibrium dynamics in non-Hermitian systems has attracted significant interest, particularly due to the skin effect and its associated anomalous phenomena. Previous studies have primarily focused on initial states with a definite particle number. Here, we present a systematic study of non-reciprocal quench dynamics in the pairing states with indefinite particle number. Our study uncovers a range of novel behaviors. Firstly, we demonstrate a universal tendency towards half-filling of particle density at late times. At early times for certain initial states, we observe a chiral wavefront in both particle number distribution and charge inflow, associated with a sharp decrease in particle number. Furthermore, we find that non-Hermiticity could enhance the growth of entanglement in the initial stages of evolution. In the intermediate time regime, the characteristic skin effect leads to particle accumulation on one side, leading to a pronounced reduction in entanglement entropy. Moreover, our results reveal the presence of the quantum Mpemba effect during the restoration of U(1) symmetry. Our findings open new avenues for exploring exotic dynamic phenomena in quantum many-body systems arising from the interplay of symmetry breaking and non-Hermiticity.
\end{abstract}

\maketitle

{\em Introduction.}---The study of non-equilibrium dynamics, especially in open systems, has recently emerged as a major research frontier, revealing intriguing phenomena beyond the scope of conventional equilibrium descriptions~\cite{Zoller2008NP,Zoller2011NP,Ott2013PRL,Wunner2014PRA,Liu2018PRL,Luo2019NC,AndreasPRB,Bryce2019NJP,Ashida2020PRB,Altman2020PRL,Liu2020PRL,Takahashi2020PTEP,Huse2020PRX,ALtman2020PRB,Grover2021RXQuantum,Hsieh2021PRR,MarcoPRB2021,Thoss2021PRB,Wald2021IOP,David2021PRR,Yi2021PRR,SchiroPRB2023,KawabataPRX2023,moreira2024arxiv,paz2024arxiv,Wang2025PRXQuantum,despres2025arxiv,gao2025arxiv,glatthard2025arxiv,Capecelatro2025arxiv}. Non-Hermitian Hamiltonians offer an effective theoretical framework for describing open quantum systems~\cite{Ashida2020Advance,Prosen2008NJP,Cooper2020PRL,Prosen2020JSM,Cui2020PRA,Yu2022PRA,Gyu2022NP}. Such non-Hermiticity naturally emerges when a system couples to an environment, introducing dissipation, gain, or non-reciprocal interactions. Moreover, non-Hermitian Hamiltonians provide a conceptually valuable description for quasiparticles possessing finite lifetimes due to mechanisms like electron-phonon coupling or disorder scattering\cite{Fu2018PRL,Fu2020PRL,Yang2025arXiv}. The investigation of non-Hermitian systems has unveiled a wealth of exotic phenomena lacking counterparts in Hermitian systems~\cite{Levitov2009PRL,Taylor2011PRB,Mahito2011PRB,Heiss2012IOP,Nori2017PRL,Hubert2017PRL,Ueda2018PRL,Gong2018PRX,Clerk2018PRX,Xiong2018IOP,Ganainy2018NP,Torres2018EPJ,Sato2019PRL,Nori2019PRL,Ashvin2019PRL,Kawakami2019PRL,Zhai2020NP,Yin2020PRB,Torres2020JPM,Iskin2021PRA,Zhou2022NC,Shinsei2021PRL,Yin2022PRB,Yan2022PRL,Hu2024PRL,Thomale2019PRB,Gong2019PRL,Shinsei2020PRR,Li2020NC,Bergholtz2021RMP,Chen2021SB,Ding2022Review,Fu2024PhysRevB,Yang2023PRR,Yang2023PRL,Sato2023Review,Hu2024CPL,Hu2024arxiv,Kou2020PRB,Kou2020arxiv,Sun2020PRR,Hu2021PRL,Hu2023NC,Chen2024PRL,Chen2023PRL,Ren2024NCP,Ren2022APL,Ye2022PRB,Ye2023PRB,Ma2022Nature,Ma2022NP,Zhang2024PRB,Zhang2023PRB,Li2024PRL,Roy2024JHEP,Roy2024NC,Zhu2024PRL,Liu2025arXiv,Kunst2021RMP,Chan2014PRX,Ueda2017NC,KawabataPRL2017} such as the celebrated non-Hermitian skin effect (NHSE) with macroscopic accumulation of eigenstates localized at the system's edges~\cite{Wang2018PRL,Murakami2019PRL,Wang2018PRL2,Torres2018PRB,Masatoshi2019PRX,Murakami2019PRL,Sato2020PRL,FangPRL2020,Wang1029PRL,Zhang2021NC,Xue2020NP,Thomale2020NP,Ken2020PRB,Zhao2022PRR,Longhi2021PRB,Park2021PRB,Shuichi2021PRB,Zhang2021PRB,Fang2022NCskineffect,2022PRAL061302,feng2025NumericalInstability,Sun2021PRL,Zhang2023PRL,Zhang2024PRL,Yang2024arXiv,Wang2024PRX,Sun2024PRB,Hu2025SB,Guo2025PRB,Manna2023NPCP,Borgnia2020PRL}.

While significant progress has been made in understanding non-Hermitian dynamics, much of the focus has been on systems with particle-number conservation~\cite{Wang1029PRL,Xiao2020NP,Zhang2021NC,
Hong2020PhysRevB,Yi2021PRR,SchiroPRB2023,Hu2024PRL,Levitov2009PRL,
Yan2022PRL,Xue2020NP,Deng2025PRB,KawabataPRX2023,PhysRevLett.120.185301}. However, many physical processes particularly those involving Cooper pairing and superconductivity, inherently lead to states that lack a definite particle number, breaking U(1) symmetry. The non-equilibrium dynamics of such U(1)-breaking states under non-Hermitian evolution, and the interplay between particle number fluctuations and unique non-Hermitian features like the NHSE, remain largely unexplored. This regime presents fundamental challenges and opportunities to discover novel phenomena arising from the confluence of non-unitary dynamics, non-reciprocity, and symmetry breaking.

Probing the non-equilibrium evolution of quantum many-body states requires analyzing multiple facets of their properties. Fundamental quantities like particle density provide essential information about the spatial distribution and transport of particles, crucial for understanding phenomena like NHSE. Beyond density,
quantum entanglement encapsulates the unique non-classical correlations fundamental to quantum systems~\cite{RevModPhys.82.277,Calabrese2004IOP,Calabrese2017PNAS, PhysRevLett.120.185301,Nahum2019PRX,Bardarson2019SPP,Andrew2020PRR,Dora2021PRB,Imura2022PRB,Dora2022PRL,SchiroPRB2023,KawabataPRX2023,ImuraPRB2023,Xu2023arxiv,Li2024PRB,Tonielli2020PhysRevLett}. Monitoring its evolution, often quantified as entanglement entropy (EE), provides a valuable dynamic window into a system's properties~\cite{Luca2019SPP,Ashida2020PRB,Diehl2021PRL,Nahum2017PhysRevX,
MarcoPRB2021,Romito2023SPP,Fang2024PRB,Jian2020PhysRevB}. Furthermore, to specifically track the restoration of broken U(1) symmetry, entanglement asymmetry (EA) has been introduced as a sensitive probe derived from entanglement, capable of detecting subtle changes in the symmetry properties of the state ~\cite{AresNC2022}.  Analyzing EA dynamics in Hermitian systems has revealed intriguing symmetry restoration phenomena, including the quantum Mpemba effect (QME)~\cite{PhysRevB.100.125102,PhysRevLett.127.060401,Zhang2025NC,PhysRevLett.133.140404,PhysRevE.111.014133,PhysRevLett.131.080402,PhysRevA.110.022213,PhysRevLett.133.010403,PhysRevResearch.3.043108,PhysRevA.106.012207,PhysRevE.108.014130,PhysRevResearch.5.043036,PhysRevResearch.6.033330,PhysRevA.110.042218,Longhi2024,PhysRevLett.133.136302,Boubakour_2025,PRXQuantum.6.010324,PhysRevB.111.104506,Longhi2025mpembaeffectsuper,PhysRevA.111.022215,Graf_2025,Van2025PRL,Vucelja2025arXiv,Zoller2024PRL,Li2024ImaginaryMpemba,CaceffoIOP2024,Clark2024arXiv}, where states initially exhibiting stronger symmetry breaking can paradoxically restore symmetry faster~\cite{AresNC2022,AresSciPostPhys2023,MurcianoIOP2024,Chalas2024IOP,yamashika2024arxiv,Rylands2024IOP,Yamashika2024PRB,Rylands2024PRL,PhysRevLett.133.140405,Yao2024mpembalocalization,Zhang2025BreakingDynamic, digiulio2025Mpemba,PhysRevB.111.L140304,PhysRevB.111.104312,Ferro_2024,Klobas_2024}. While these quantities have been extensively studied in various contexts, their interplay and behavior under non-Hermitian evolution, particularly for initial states lacking particle-number conservation, remain elusive. This motivates fundamental questions: How are particle density profiles, entanglement growth, and U(1) symmetry restoration affected by non-Hermitian dynamics and the NHSE in states without definite particle number?

To address these essential questions, we investigate the non-equilibrium dynamics of a superconducting pairing state under the time evolution of a typical non-Hermitian Hamiltonian, termed the Hatano-Nelson (HN) model~\cite{HN1997PRB}, which involves non-reciprocal hopping of fermions exhibiting the skin effect. This choice of a U(1) symmetry-breaking initial state marks a significant departure from the aforementioned studies that focused on number-conserving initial states. Our approach therefore provides a direct window into the unique phenomena emerging from the interplay between initial particle number fluctuations and non-Hermitian dynamics. For the time evolution of the pairing state under HN model, the total particle number is not conserved. Here, we demonstrate a universal late time convergence to half-filling, arising from the skin effect, regardless of the initial particle number. Furthermore, we uncover a range of novel phenomena in the short-time dynamics arising from the interplay of U(1) symmetry breaking and non-Hermiticity. (1) We observe a chiral wavefront in the particle density dynamics, which is also manifested in the behavior of particle inflow due to non-Hermiticity for certain initial states. (2) The rate of entanglement growth is enhanced at early times while greatly suppressed at late times by the non-Hermiticity. (3) In the process of U(1) symmetry restoration, QME persists with the introduction of non-Hermiticity. Crucially, these universal results are robust across a wide range of parameters within the model, holding significant promise for experimental verification.

{\em Model and Method.}---We employ the initial state with charge fluctuation by considering the ground state of the one-dimensional fermionic Hamiltonian involving Cooper pairing:
\begin{equation}
\begin{aligned}
\hat{H}_{\text{ini}}=-\sum_{j}(J\hat{c}_{j}^{\dagger}\hat{c}_{j+1}+\Delta \hat{c}_{j}^{\dagger}\hat{c}_{j+1}^{\dagger}+\text{h.c.})
-\mu\sum_{j}\hat{c}_{j}^{\dagger}\hat{c}_{j}.
 	\label{Eq1}
\end{aligned}
\end{equation}
where $\hat{c}_{j}^{\dagger}$ ($\hat{c}_{j}$) represent the creation (annihilation) operators, $J$ is the amplitude of nearest-neighbor (NN) hopping, $\mu$ is chemical potential, and $\Delta$ denotes the pairing amplitude. The initial state is thus  a superconducting BCS pairing state breaking charge U(1) symmetry. Through Jordan-Wigner transformation (see the section~I of the supplemrental material (SM)~\cite{Ref_SM_V2} for the details), the fermionic Hamiltonian in \Eq{Eq1} is straightforwardly transformed to a spin-1/2 chain as follows:
\begin{equation}
\begin{aligned}
\hat{H}_{\text{ini}}=-&\frac{J+\Delta}{2}\sum_{j}\hat{\sigma}_{j}^{x}\hat{\sigma}_{j+1}^{x}
-\frac{J-\Delta}{2}\sum_{j}\hat{\sigma}_{j}^{y}\hat{\sigma}_{j+1}^{y}\\
-&\frac{\mu}{2}\sum_{j}\hat{\sigma}_{j}^{z}.
 	\label{Eq2}
\end{aligned}
\end{equation}
$\Delta$ breaks U(1) spin rotational symmetry around $z$-axis. We implement anti-periodic boundary condition (APBC) of \Eq{Eq1}, corresponding the periodic boundary condition (PBC) of spin-1/2 chain in \Eq{Eq2} under Jordan-Wigner transformation for even number of lattice sites.
The ground state of \Eq{Eq1} is conveniently expressed in the spin representation as a superposition of tilted ferromagnetic states~\cite{AresNC2022}:
\begin{equation}
\begin{aligned}
| \psi(0) \rangle= &\frac{1}{\sqrt{2}}(| F,\theta \rangle + | F,-\theta \rangle),\\
| F,\theta \rangle=&\underset{j}{\otimes} (\cos\frac{\theta}{2}|\uparrow_{j}  \rangle+\sin\frac{\theta}{2} | \downarrow_{j}\rangle),
 	\label{Eq3}
\end{aligned}
\end{equation}
where $\theta$ represents the tilted angle characterizing the strength of U(1) symmetry breaking. The relations between $\theta$, $\Delta$ and $\mu$ satisfy~\cite{AresNC2022,AresSciPostPhys2023,ShrockPhysRevB}
\begin{equation}
\begin{aligned}
\text{cos}^{2}\theta=\frac{1-\Delta}{1+\Delta},
\Delta^2+(\mu/2)^2=1.
 	\label{Eq4}
\end{aligned}
\end{equation}
This relation is not an arbitrary constraint but is the necessary condition for our initial state to be a simple, product-form ground state of a local XY Hamiltonian~\cite{ShrockPhysRevB}. This allows us to parameterize the initial degree of U(1) symmetry breaking with the single angle $\theta$, a method also employed in Refs.~\cite{AresNC2022,AresSciPostPhys2023} to study Hermitian quench dynamics from such states. To investigate the non-Hermitian dynamics, we employ the HN model~\cite{HN1997PRB} whose Hamiltonian is:
\begin{equation}
\begin{aligned}
\hat{H}_{\text{evo}}=-\sum_{j}(t_{L}\hat{c}_{j}^{\dagger}\hat{c}_{j+1}+t_{R}\hat{c}_{j+1}^{\dagger}\hat{c}_{j}),
 	\label{Eq6}
\end{aligned}
\end{equation}
where $t_{L}=J+\gamma$, $t_{R}=J-\gamma$ are the non-reciprocal hopping coefficients.  Crucially, the non-reciprocity ($\gamma \ne 0$) inherent in these hopping coefficients is responsible for the skin effect under open boundary conditions (OBC).

For non-unitary time evolution, the normalized state $|\psi(t)\rangle$ at time $t$ is given by:
\begin{equation}
\begin{aligned}
 | \psi(t)  \rangle =\frac{e^{-i\hat{H}_{\text{evo}}t}| \psi(0) \rangle}{\sqrt{  \langle \psi(0) |e^{i\hat{H}_{\text{evo}}^{\dagger}t}e^{-i\hat{H}_{\text{evo}}t} | \psi(0)  \rangle }}.
 	\label{Eq7}
\end{aligned}
\end{equation}
For numerical calculations, we use the quantum software {\sf TensorCircuit-NG}~\cite{*[{ }] [{. https://github.com/tensorcircuit/tensorcircuit-ng.}] Zhang2022_z}. The energy unit is set as $J=1$, system size $L=64$ and OBC are employed. Without loss of generality, we focus on $\gamma \geq 0$, which results in the skin effect where particles accumulate at the left boundary. It is important to note that the combination of non-reciprocity and the large system size can introduce significant numerical errors in the direct calculation of $e^{-i\hat{H}_{\text{evo}}t}$ \cite{feng2025NumericalInstability}. To maintain numerical accuracy, we utilize a small time-step decomposition of the evolution and perform calculations with a high precision of 250 digits (See the details in section~II of the SM~\cite{Ref_SM_V2}).

\begin{figure}[t]
	\centerline{\includegraphics[scale=0.30]{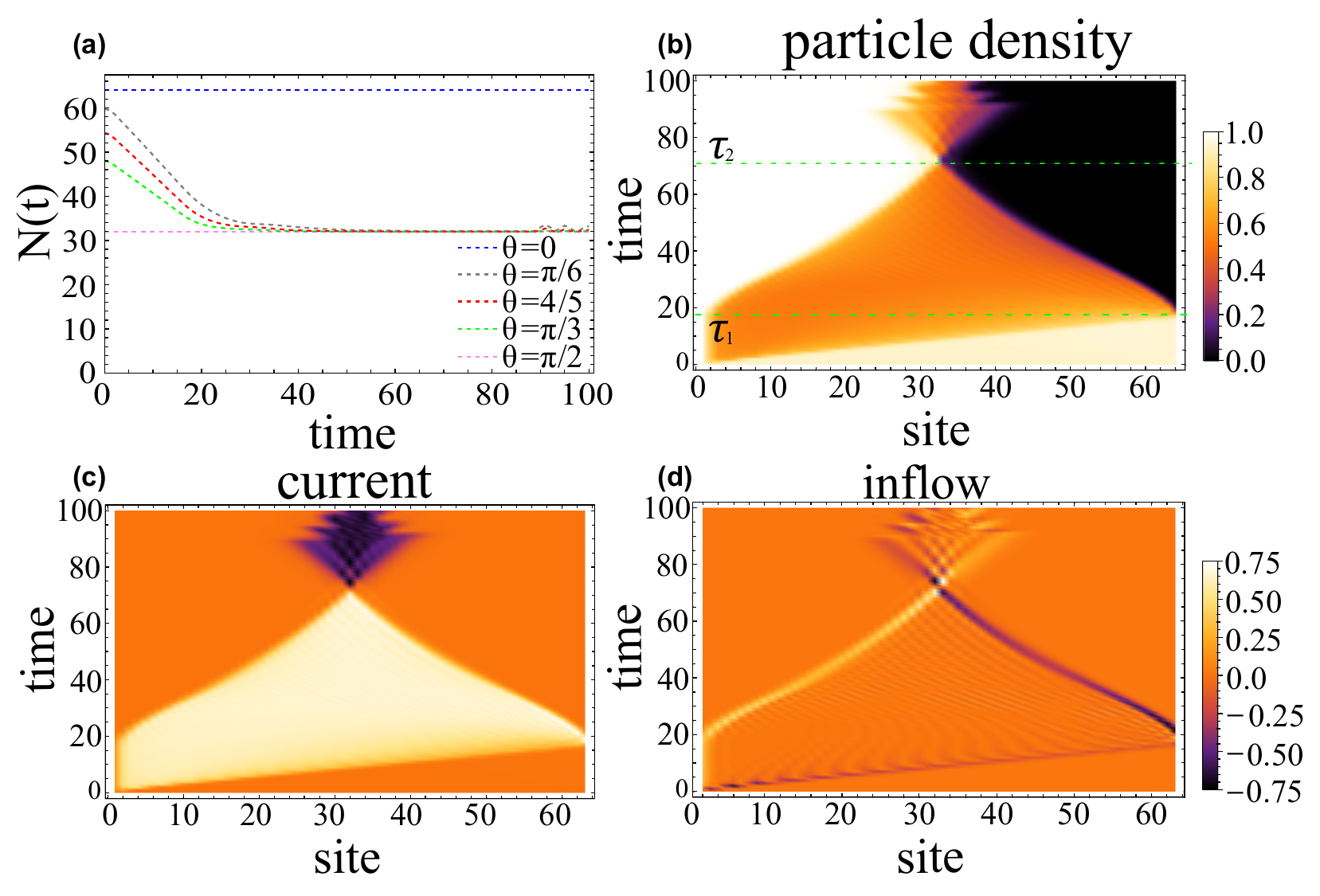}}
	\caption{(a) Evolution of the total particle number starting from various initial tilted ferromagnetic states.  (b) Dynamics of local particle density $n_{j}$. Characteristic timescales $\tau_{1}$ and $\tau_{2}$ are marked by horizontal dashed lines. Dynamics of (c) current density $I_{j}$, and (d) particle inflow rate $\sigma_{j}$.  In (c), positive current indicates flow from left to right, and negative current from right to left. In (d), positive inflow represents particle gain, and negative inflow represents particle loss. For (b)-(d), the initial state is the tilted ferromagnetic state with $\theta=\pi/6$. The system size is $L=64$ and the non-reciprocity strength is $\gamma=0.8$ for all figures.}
	\label{Fig1}
\end{figure}

\begin{figure*}[tb]
	\centerline{\includegraphics[scale=0.35]{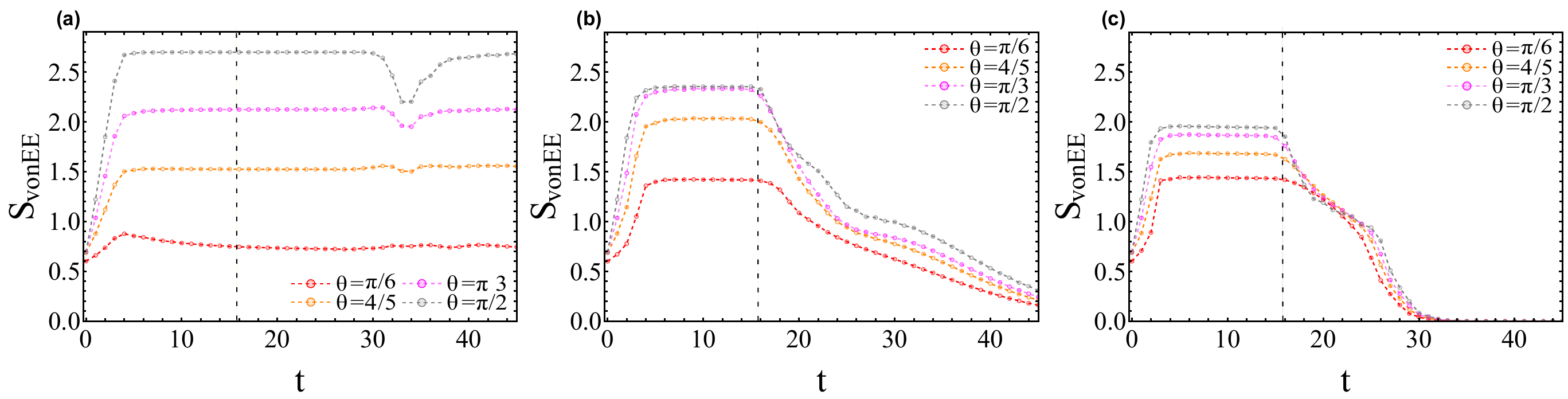}}
	\caption{Entanglement entropy under open boundary conditions. (a), (b) and (c) correspond to $\gamma =0.0$, 0.2 and 0.6, respectively, where the legends represent different tilted angles. The traced subsystem $B$ with length $l=6$ is located at the left end of the system. The total system size is $L=64$ and the characteristic timescale $\tau_{1}$ is marked by vertical dashed line.}
	\label{Fig2}
\end{figure*}

{\em Universal late-time half-filling convergence.}
Different from the initial state with particle-number conservation, a prominent feature of the U(1) symmetry breaking state under time evolution of HN model is that the particle number is not conserved. We first examine the dynamics of the total particle number, defined by
\begin{equation}
\begin{aligned}
\overline{N}(t)=\sum_{j} \langle \psi (t)| \hat{n}_{j}|\psi (t)  \rangle,
 	\label{Eq8-1}
\end{aligned}
\end{equation}
where $\hat{n}_{j}=\hat{c}^{\dagger}_{j}\hat{c}_{j}$ is the on-site particle density. The results for $\overline{N}(t)$, exemplified in Fig. 1(a) for $\gamma=0.8$, reveal a universal long-time behavior. Here, ``universal" refers to the robust relaxation towards half-filling ($L/2$) for any initial tilted ferromagnetic state that breaks the U(1) symmetry ($\theta > 0$) within the Hatano-Nelson model under OBC. While we do not claim this extends to all non-Hermitian systems, we expect it to be a general feature for models exhibiting a prominent non-Hermitian skin effect.

This phenomenon is rigorously proven in the strong non-reciprocity limit. As we demonstrate analytically in section III-A of the SM~\cite{Ref_SM_V2} , for $\gamma=1$, the total particle number is guaranteed to asymptotically approach $L/2$. For weaker non-reciprocity ($\gamma<1$), while a full analytical proof is more challenging, our numerical results provide strong evidence for this trend. The clear convergence for $\gamma=0.8$ in Fig.~\ref{Fig1}(a) supports this. Furthermore, simulations presented in the SM~\cite{Ref_SM_V2}  confirm that relaxation to half-filling still occurs for smaller $\gamma$, even under different boundary conditions for state preparation and evolution, showcasing the robustness of this phenomenon. This suggests that $\gamma$ primarily controls the rate of convergence to this universal steady state. In stark contrast, the number-conserving ferromagnetic state ($\theta=0$), being an eigenstate of the local density operator, remains fully occupied throughout the evolution. We also note that for the case of PBC, which is discussed in the SM~\cite{Ref_SM_V2} , the total particle number for a generic initial state also tends towards $L/2$ in the steady state.

{\em Density dynamics.}--- To further explore the dynamical characteristics reflected in the particle number, we now examine the time evolution of the on-site particle density $\langle \hat{n}_{j}\rangle$, exemplified in Fig.~\ref{Fig1}(b) for $\theta=\pi/6$~\cite{Ref_SM_V2}. While the initial state exhibits a uniform particle density, distinct behaviors emerge at different timescales.  At long times, there is a clear spatial separation of particles due to the skin effect. The particle density becomes concentrated on the left side of the system, leaving the opposite side sparsely populated. Intriguingly, short-time dynamics reveal a contrasting exotic behavior.  Instead of accumulation, we observe an initial reduction of particle density near the left edge. This sharp depletion is linked to a chiral wavefront propagating with twice the maximum group velocity of quasiparticles.  Specifically, particle density begins to decrease as the wavefront reaches a given site. Here, the group velocity is defined as $v_{\text{g}}=\frac{\mathrm{d} }{\mathrm{d} k}\text{Re}[E(k)] $, where $E(k)=-2J\text{cos}k-2i\gamma \text{sin}k$ is the energy spectrum of the HN model under PBC~\cite{ModakIOP2020,ImuraPRB2023,CevolaniPRB2018}.

With increasing time, two characteristic timescales, $\tau_{1}$ and $\tau_{2}$, become evident.  The first $\tau_{1}$ corresponds to the arrival time of the chiral wavefront at the right boundary, at which point the total particle number approaches $L/2$ for large $\gamma$. Around this time, the particle density begins to change from both boundaries inwards: increasing from the left edge and decreasing from the right edge towards the center. Upon reaching the second timescale, $\tau_{2}$, the density distribution stabilizes into a steady state, with only minor oscillations remaining in the central region.  This steady state distribution clearly demonstrates the skin effect, with the left half of the system almost fully occupied and the right half nearly empty. In stark contrast, an initial tilted antiferromagnetic (AFM) state lacks this chiral wavefront; its dynamics immediately resemble the post-$\tau_{2}$ behavior of the tilted ferromagnetic state, likely because the AFM state starts at half-filling.
Detailed results are presented in section~V of the SM~\cite{Ref_SM_V2}.

The time-evolution of particle density is determined by the continuity equation as:
\begin{equation}
\begin{aligned}
\frac{\partial n_{j}}{\partial t}+(I_{j}-I_{j-1})=\sigma_{j},
 	\label{Eq8-2}
\end{aligned}
\end{equation}
where $I_j=iJ(\hat{c}^{\dagger}_{j+1}\hat{c}_{j}-\hat{c}^{\dagger}_{j}\hat{c}_{j+1})$ represents the current flowing out of site $j$ and into site $j+1$, and $\sigma_j$ denotes the inflow of particles into site $j$ from the external environment.  In Hermitian systems without particle dissipation, $\sigma_j$ vanishes, and \Eq{Eq8-2} reduces to the standard continuity equation. To gain further insight into the particle dynamics, particularly the influence of current density and particle inflow, we show the dynamical results for current density and particle inflow in Fig.~\ref{Fig1}(c) and (d), respectively.

Similar to the particle density, characteristic timescales $\tau_1$ and $\tau_2$ are also evident in the dynamics of both current density and particle inflow.  In the short-time regime ($\tau < \tau_1$), a chiral wavefront emerges in both the current density and particle inflow profiles. The characteristic time $\tau_1$ corresponds to the moment when this chiral wavefront reaches the right boundary of the system. Notably, the wavefront in the particle inflow in Fig.~\ref{Fig1}(d) represents a significant particle sink, indicating a substantial loss of particle number at given space-time location. The  wavefront of current in Fig.~\ref{Fig1}(c) appears as a sharp spatial gradient of current density. Although the negative current gradient implies particle accumulation according to the continuity equation \Eq{Eq8-2}, the observed net effect at the wavefront is particle depletion. Thus, the particle number reduction associated with the wavefront in Fig.~\ref{Fig1}(b) arises from the effect of particle inflow acting as a propagating sink, which dominates over the particle-accumulating effect of the current induced by non-reciprocity. In the intermediate stage ($\tau_1 < \tau < \tau_2$), the rapid increase in particle number at the left edge and decrease at the right edge are also driven by the particle inflow, as visualized in Fig.~\ref{Fig1}(d).  For late times ($\tau > \tau_2$), both the current and particle inflow diminish, exhibiting only small, time-decaying oscillations.  Results for other values of $\gamma$ show qualitatively consistent behaviors, as included in
section~V of the SM~\cite{Ref_SM_V2}. Moreover, our analysis of the density dynamics under APBC, detailed in the the section~VI of the SM~\cite{Ref_SM_V2}, yields qualitatively distinct conclusions, further demonstrating that the anomalous phenomena observed under OBC are attributed to the skin effect.

{\em Entanglement dynamics.}---We now turn to the dynamics of bipartite entanglement, where we employ the  bipartite von Neumann EE, defined as:
\begin{equation}
\begin{aligned}
S_{\text{vonEE}}=-\text{Tr}_{A}[\rho_{A}\text{ln}\rho_{A}].
 	\label{Eq12}
\end{aligned}
\end{equation}
Here, the system is bipartitioned into subsystems $A$ and $B$, and $\rho_{A}=\text{Tr}_{B}[\rho]$ is the reduced density matrix of subsystem $A$. For the Hermitian case (Fig.~\ref{Fig2}(a)), the EE for different initial tilted states initially increases and subsequently saturates to a steady-state value. In contrast, the non-Hermitian dynamics, shown in Fig.~\ref{Fig2}(b) and (c), exhibit more complex phenomena. Similar to the particle density dynamics, the entanglement evolution displays distinct behaviors in different temporal regimes. In the stage $\tau < \tau_1$, EE increases with evolution time and subsequently saturates, qualitatively consistent with the conventional behavior of entanglement dynamics in Hermitian systems. This trend is qualitatively captured by the quasiparticle picture, as demonstrated in section~IV of the SM~\cite{Ref_SM_V2}.
An intriguing observation, illustrated in Fig.~\ref{Fig2}(b)-(c), is that non-Hermiticity can enhance both the growth rate and the maximum value of EE during time evolution for specific initial states and small $\gamma$.

In contrast to Hermitian systems, the EE exhibits a rapid decrease in the temporal regime $t > \tau_1$, a behavior fundamentally different from the Hermitian case. This is directly attributed to the particle density dynamics driven by the skin effect. As shown in Fig.~\ref{Fig1}(b), the skin effect leads to particle accumulation on the left side, with the occupied region expanding over time.  Consequently, near the boundary between subsystems A and B, the system increasingly resembles a product state, which inherently suppresses entanglement. Thus, the observed EE reduction is a direct consequence of the particle number distribution arising from the skin effect.

Given the skin effect, we anticipate that the EE depends on the bipartition subsystem's location. To investigate this, we calculate the EE for different subsystem positions, with detailed results presented in section~V of the SM~\cite{Ref_SM_V2}.  When the small subsystem is located centrally, the qualitative behavior of EE dynamics remains similar to the cases where the subsystem is at the left or right boundary. However, the reduction in EE after $t > \tau_1$ is significantly less pronounced compared to the results in Fig.~\ref{Fig2}(b) and (c). This milder reduction can be naturally understood through the particle number distribution induced by the skin effect, as the particle density fluctuation is stronger at the middle region.

\begin{figure*}[tb]
	\centerline{\includegraphics[scale=0.4]{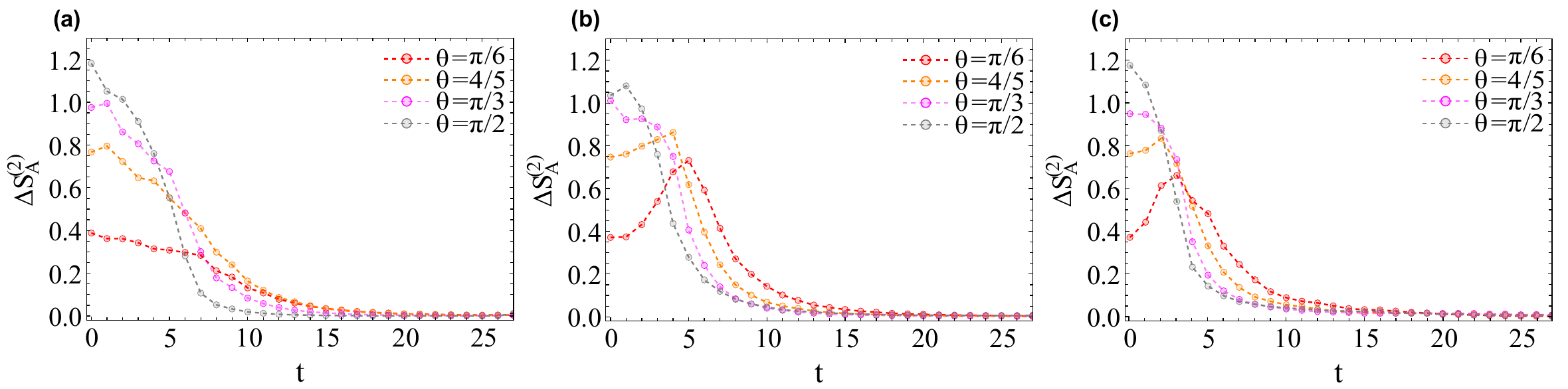}}
	\caption{Entanglement asymmetry under the open boundary conditions. (a), (b) and (c) correspond to $\gamma=0.0$, $0.2$ and $0.6$, respectively, where the legends represent different tilted angles. The traced subsystem $B$ with length $l=12$ is located at the left end of the system. The total system size is $L=64$.}
	\label{Fig3}
\end{figure*}

{\em Quantum Mpemba effect.}---Recently, researchers have found that EE can also be used to quantitatively characterize the degree of broken U(1) symmetry of the subsystem $A$, generated by the particle number operator $\hat{N}=\sum_{j}\hat{c}_{j}^{\dagger}\hat{c}_{j}$. Ref.~\cite{AresNC2022} defines R$\acute{e}$nyi EA to describe the symmetry breaking, which is given by
\begin{equation}
\begin{aligned}
\Delta S^{(n)}_{A}= S_{n}(\rho_{A,N})-S_{n}(\rho_{A}),
 	\label{Eq13}
\end{aligned}
\end{equation}
where $S_{n}(\rho)$ is $n$-th R$\acute{e}$nyi EE, $\rho_{A,N}=\sum_{q\in \mathbb{Z} }\Pi_{q}\rho_{A}\Pi_{q}$ and $\Pi_{q}$ is the projector onto the eigenspace of $N_{A}$ with particle $q\in \mathbb{Z}$. The magnitude of EA is proportional to the degree of U(1) symmetry breaking, with the limit $\Delta S^{(n)}_{A}=0$ corresponding to an unbroken U(1) symmetry.

Recent research has revealed a novel and universal QME in dynamic symmetry restoration, where systems with greater initial symmetry breaking recover symmetry faster~\cite{AresNC2022,MurcianoIOP2024,CaceffoIOP2024,digiulio2025Mpemba,AresSciPostPhys2023,Chalas2024IOP,yamashika2024arxiv,Rylands2024IOP,Yamashika2024PRB,Rylands2024PRL,PhysRevLett.133.140405,Yao2024mpembalocalization,Zhang2025BreakingDynamic,Li2024ImaginaryMpemba}.  A key question is whether symmetry restoration and the QME can emerge in non-Hermitian systems.  Fig.~\ref{Fig3} presents the second-order Rényi EA for varying non-Hermiticity strengths ($\gamma$).  Consistent with prior work~\cite{AresNC2022}, \Fig{Fig3}(a) shows a clear curve crossing for different tilted angles ($\theta$) in the Hermitian case, confirming the QME. Notably, a pronounced QME also exists in the non-Hermitian scenario, as evidenced by curve crossings in \Fig{Fig3}(b) and (c). Intriguingly, this QME appears enhanced compared to the Hermitian case, with curve crossings occurring earlier in time. This enhancement is accompanied by non-monotonic behavior in the early-time EA evolution for some initial states, where EA initially increases before decreasing towards symmetry restoration. Further analysis reveals that the strength of this enhancement is itself non-monotonic with respect to the non-reciprocity strength $\gamma$. Our results suggest the existence of a critical value of $\gamma$ where the QME is most pronounced, before the effect weakens again at stronger non-reciprocity. A detailed discussion of this phenomenon is provided in the section V-E of the SM~\cite{Ref_SM_V2}. At late times, the EA results show a restoration of U(1) symmetry, mirroring the behavior observed in Hermitian systems.

{\em Conclusions and discussions.}---In this Letter, we investigate the quench dynamics of the pairing state with indefinite particle number under the time evolution of HN model. For the first time, we demonstrate the emergence of universal half-filling behavior in the late-time limit of the HN model. Beyond this, we unveil a rich array of novel phenomena stemming from the interplay between U(1) symmetry breaking and non-Hermiticity. Specifically, we observe a chiral wavefront in the particle density dynamics, non-Hermiticity enhanced entanglement growth and QME. Our findings open new routes for understanding quantum many-body dynamics in the presence of both symmetry breaking and dissipation.

The evolution Hamiltonian can be derived as an effective Hamiltonian from a Lindblad master equation~\cite{Fang2024PRB}, i.e., $\hat{H}_{\text{eff}}=\hat{H}-2i\gamma \sum_{j}\hat{L}_{j}^{\dagger}\hat{L}_{j}$, where $\hat{H}=-J \sum_{j}(\hat{c}_{j}^{\dagger}\hat{c}_{j+1}+\text{h.c.})$ and $\hat{L}_{j}=(\hat{c}_{j}+i\hat{c}_{j+1})/\sqrt{2}$ is Lindblad operator describing quantum jumps. The Lindblad master equation describes the coupling between the subsystem and the external environment, allowing for the dissipation of energy and particles. It is important to clarify the physical context in which the non-unitary dynamics governed by Eq.~(6) is valid. The normalized evolution prescribed by Eq.~(6) does not represent the full ensemble dynamics of the open system; rather, it describes the system's evolution along a specific ``quantum trajectory" conditioned on the absence of any quantum jumps~\cite{Carmichael1993PhysRevLett,Knight1998RevModPhys}. This post-selection scheme is a standard and powerful paradigm for exploring the intrinsic effects of non-Hermiticity, such as the skin effect, which are the primary focus of our work.

A promising experimental platform for realizing this system lies in ultracold atomic systems, where dissipation can be engineered via coupling the subsystem to a reservoir~\cite{Yan2020PRL,Zoller2011NP} or utilizing controlled radiation processes between different hyperfine states~\cite{Jo2022NP,Yan2022PRL}. In such platforms, realizing the post-selected dynamics requires continuous monitoring of the system to detect quantum jumps~\cite{Vijay2011PhysRevLett,Minev2019Nature}. By post-selecting the experimental runs where no jumps were detected, one can reconstruct the conditional dynamics described by Eq.~(6). Furthermore, quench dynamics can be achieved by sudden changes in system parameters~\cite{Zwerger2008RMP}. Subsequently, atomic densities can be probed with high precision using time-of-flight measurements~\cite{Svistunov2008PRL,Bloch2005PRL} and in situ imaging techniques~\cite{Takahashi2016NP,Kuhr2010Nature}. Given the well-established experimental techniques in ultracold atomic systems, this platform could facilitate further verification and understanding of the anomalous phenomena we report.

The numerical data that support the findings of this study are available in the Figshare repository~\cite{GuoData2024}.

{\em Acknowledgments.}----S.G. and Z-.X.L. are supported by Beijing Natural Science Foundation (No. JR25007) and the National Natural Science Foundation of China (No. 12347107 and No. 12474146). S.Y. is supported by the National Natural Science Foundation of China (No. 12222515), the Science and Technology Projects in Guangdong Province (No. 2021QN02X561), and the Science and Technology Projects in Guangzhou City (No. 2025A04J5408). S.-X.Z. acknowledges the support from Innovation Program for Quantum Science and Technology (2024ZD0301700).

\bibliography{ref}

\onecolumngrid
\newpage


\section*{Supplementary material}

# Supplemental material for
# "Skin Effect Induced Anomalous Dynamics from Charge-Fluctuating Initial States"

Sibo Guo[1,2], Shuai Yin[3,4], Shi-Xin Zhang[1,*] and Zi-Xiang Li[1,2†]

[1]*Beijing National Laboratory for Condensed Matter Physics & Institute of Physics,
Chinese Academy of Sciences, Beijing 100190, China*
[2]*University of Chinese Academy of Sciences, Beijing 100049, China*
[3]*Guangdong Provincial Key Laboratory of Magnetoelectric Physics and Devices,
School of Physics, Sun Yat-Sen University, Guangzhou 510275, China and*
[4]*School of Physics, Sun Yat-Sen University, Guangzhou 510275, China*

(Dated: April 30, 2025)

This supplemental material provides further details and supporting evidence for the results presented in the main text, organized into six sections. Section I details the derivation of Eq. (1) from Eq. (2) in the main text using the Jordan-Wigner transformation. Section II outlines the numerical implementation for simulating fermionic Gaussian states. Subsequently, section III presents a quantitative analysis supporting the arguments concerning the long-time particle density dynamics under both open (OBC) and anti-periodic (APBC) boundary conditions. A more detailed description of the non-Hermitian quasiparticle picture is discussed in section IV. Section V is dedicated to additional results under OBC, comprising five subsections: sections V-A to V-C provide detailed particle density dynamics for various tilted angles ($\theta$) and non-reciprocity values ($\gamma$), including tilted ferromagnetic and antiferromagnetic states; sections V-D and V-E explore entanglement dynamics, examining the dependence of entanglement entropy (EE) and entanglement asymmetry (EA) on subsystem position, the enhancement of EE growth rates, and the non-monotonicity in EA for different $\gamma$. Finally, section VI investigates density and entanglement dynamics under APBC. Throughout this supplement, a consistent system size of $L = 64$ is maintained.

## SECTION I. JORDAN-WIGNER TRANSFORMATION

In this section, we aim to derive Eq. (1) in the main text from Eq. (2) using the Jordan-Wigner transformation. This transformation is a powerful tool for mapping spin operators to fermionic operators in one-dimensional systems. The initial Hamiltonian in Eq. (2) is

$$\hat{H}_{\text{ini}} = -\frac{J+\Delta}{2}\sum_j \hat{\sigma}_j^x \hat{\sigma}_{j+1}^x - \frac{J-\Delta}{2}\sum_j \hat{\sigma}_j^y \hat{\sigma}_{j+1}^y - \frac{\mu}{2}\sum_j \hat{\sigma}_j^z. \tag{1}$$

The Jordan-Wigner transformation between spin operators and fermionic operators is:

$$\begin{aligned}\hat{S}_j^+ &= e^{i\pi \sum_{m=1}^{j-1} \hat{c}_m^\dagger \hat{c}_m} \hat{c}_j^\dagger, \\ \hat{S}_j^- &= e^{-i\pi \sum_{m=1}^{j-1} \hat{c}_m^\dagger \hat{c}_m} \hat{c}_j,\end{aligned} \tag{2}$$

where $\hat{c}_j^\dagger (\hat{c}_j)$ is the fermionic annihilation (creation) operator acting on site $j$, and $\hat{S}_j^+ (\hat{S}_j^-)$ is the raising (lowering) operator for the spin at site $j$ in the spin-1/2 system. The exponential term can be simplified as:

$$e^{i\pi \hat{c}_m^\dagger \hat{c}_m} = (-1)^{\hat{c}_m^\dagger \hat{c}_m} = 1 - 2\hat{c}_m^\dagger \hat{c}_m. \tag{3}$$

Thus we have

$$e^{i\pi \hat{c}_m^\dagger \hat{c}_m} = e^{-i\pi \hat{c}_m^\dagger \hat{c}_m}. \tag{4}$$

Introducing the Pauli matrix $\hat{S}_j^\alpha = \frac{1}{2}\hat{\sigma}_j^\alpha$, where $\alpha = x, y, z$, we obtain the following relations

$$\begin{aligned}\hat{\sigma}_j^x &= \frac{1}{2}(\hat{\sigma}_j^+ + \hat{\sigma}_j^-) = e^{i\pi \sum_{m=1}^{j-1} \hat{c}_m^\dagger \hat{c}_m}(\hat{c}_j^\dagger + \hat{c}_j), \\ \hat{\sigma}_j^y &= \frac{1}{2i}(\hat{\sigma}_j^+ - \hat{\sigma}_j^-) = -ie^{i\pi \sum_{m=1}^{j-1} \hat{c}_m^\dagger \hat{c}_m}(\hat{c}_j^\dagger - \hat{c}_j).\end{aligned} \tag{5}$$

Substituting Eq. 5 into $\hat{\sigma}_j^\beta \hat{\sigma}_{j+1}^\beta$, where $\beta = x, y$, we have

$$
\begin{aligned}
\hat{\sigma}_j^x \hat{\sigma}_{j+1}^x &= e^{i\pi \sum_{m=1}^{j-1} \hat{c}_m^\dagger \hat{c}_m}(\hat{c}_j^\dagger + \hat{c}_j) e^{i\pi \sum_{m=1}^{j-1} \hat{c}_m^\dagger \hat{c}_m} e^{i\pi \hat{c}_j^\dagger \hat{c}_j}(\hat{c}_{j+1}^\dagger + \hat{c}_{j+1}) \\
&= (\hat{c}_j^\dagger + \hat{c}_j)(1 - 2\hat{c}_j^\dagger \hat{c}_j)(\hat{c}_{j+1}^\dagger + \hat{c}_{j+1}) \\
&= (\hat{c}_j^\dagger - \hat{c}_j)(\hat{c}_{j+1}^\dagger + \hat{c}_{j+1}), \\
&= \hat{c}_j^\dagger \hat{c}_{j+1}^\dagger + \hat{c}_j^\dagger \hat{c}_{j+1} - \hat{c}_j \hat{c}_{j+1}^\dagger - \hat{c}_j \hat{c}_{j+1}
\end{aligned}
\quad (6)
$$

and

$$
\begin{aligned}
\hat{\sigma}_j^y \hat{\sigma}_{j+1}^y &= (-1)e^{i\pi \sum_{m=1}^{j-1} \hat{c}_m^\dagger \hat{c}_m}(\hat{c}_j^\dagger - \hat{c}_j) e^{i\pi \sum_{m=1}^{j-1} \hat{c}_m^\dagger \hat{c}_m} e^{i\pi \hat{c}_j^\dagger \hat{c}_j}(\hat{c}_{j+1}^\dagger - \hat{c}_{j+1}) \\
&= (-1)(\hat{c}_j^\dagger - \hat{c}_j)(1 - 2\hat{c}_j^\dagger \hat{c}_j)(\hat{c}_{j+1}^\dagger - \hat{c}_{j+1}) \\
&= (-1)(\hat{c}_j^\dagger + \hat{c}_j)(\hat{c}_{j+1}^\dagger - \hat{c}_{j+1}), \\
&= -\hat{c}_j^\dagger \hat{c}_{j+1}^\dagger + \hat{c}_j^\dagger \hat{c}_{j+1} - \hat{c}_j \hat{c}_{j+1}^\dagger + \hat{c}_j \hat{c}_{j+1}.
\end{aligned}
\quad (7)
$$

Combining Eq. 6 and Eq. 7, we obtain

$$
\frac{J+\Delta}{2} \hat{\sigma}_j^x \hat{\sigma}_{j+1}^x + \frac{J-\Delta}{2} \hat{\sigma}_j^y \hat{\sigma}_{j+1}^y = \Delta(\hat{c}_j^\dagger \hat{c}_{j+1}^\dagger + \hat{c}_{j+1} \hat{c}_j) + J(\hat{c}_j^\dagger \hat{c}_{j+1} + \hat{c}_{j+1}^\dagger \hat{c}_j). \quad (8)
$$

Now, considering the periodic boundary, the boundary terms are

$$
\begin{aligned}
\hat{\sigma}_L^x \hat{\sigma}_{L+1}^x &= e^{i\pi \sum_{m=1}^{L-1} \hat{c}_m^\dagger \hat{c}_m}(\hat{c}_L^\dagger + \hat{c}_L)(\hat{c}_1^\dagger + \hat{c}_1) \\
&= e^{i\pi \sum_{m=1}^{L} \hat{c}_m^\dagger \hat{c}_m} e^{i\pi \hat{c}_L^\dagger \hat{c}_L}(\hat{c}_L^\dagger + \hat{c}_L)(\hat{c}_1^\dagger + \hat{c}_1) \\
&= (-1)^L (1 - 2\hat{c}_L^\dagger \hat{c}_L)(\hat{c}_L^\dagger + \hat{c}_L)(\hat{c}_1^\dagger + \hat{c}_1) \\
&= (-1)^L (\hat{c}_L - \hat{c}_L^\dagger)(\hat{c}_1^\dagger + \hat{c}_1) \\
&= (-1)^L (\hat{c}_L \hat{c}_1^\dagger + \hat{c}_L \hat{c}_1 - \hat{c}_L^\dagger \hat{c}_1^\dagger - \hat{c}_L^\dagger \hat{c}_1)
\end{aligned}
\quad (9)
$$

and

$$
\begin{aligned}
\hat{\sigma}_L^y \hat{\sigma}_{L+1}^y &= (-1)e^{i\pi \sum_{m=1}^{L-1} \hat{c}_m^\dagger \hat{c}_m}(\hat{c}_L^\dagger - \hat{c}_L)(\hat{c}_1^\dagger - \hat{c}_1) \\
&= (-1)e^{i\pi \sum_{m=1}^{L} \hat{c}_m^\dagger \hat{c}_m} e^{i\pi \hat{c}_L^\dagger \hat{c}_L}(\hat{c}_L^\dagger - \hat{c}_L)(\hat{c}_1^\dagger - \hat{c}_1) \\
&= (-1)^{L+1}(1 - 2\hat{c}_L^\dagger \hat{c}_L)(\hat{c}_L^\dagger - \hat{c}_L)(\hat{c}_1^\dagger - \hat{c}_1) \\
&= (-1)^L (\hat{c}_L^\dagger + \hat{c}_L)(\hat{c}_1^\dagger - \hat{c}_1) \\
&= (-1)^L (\hat{c}_L^\dagger \hat{c}_1^\dagger - \hat{c}_L^\dagger \hat{c}_1 + \hat{c}_L \hat{c}_1^\dagger - \hat{c}_L \hat{c}_1).
\end{aligned}
\quad (10)
$$

It should be noted that our initial state is a superposition of tilted ferromagnetic states, denoted as $(|F, \theta\rangle + |F, -\theta\rangle)/\sqrt{2} = \sum_n (-1)^{n(n-1)/2}(1 + (-1)^{L-n})/(\sqrt{2})\cos^n \frac{\theta}{2} \sin^{L-n} \frac{\theta}{2} |n\rangle$, where $|n\rangle$ contains $C_L^n$ different configurations corresponding $n$ fermions occupation. Regarding the expansion coefficients for the configurations $|n\rangle$, the non-vanishing terms correspond only to those where the particle number $n$ has the same parity as the system size $L$. Consequently, this leads to $e^{i\pi \sum_{m=1}^{L} \hat{c}_m^\dagger \hat{c}_m} = (-1)^L$. Combining Eq. 9 and Eq. 10, we have

$$
\frac{J+\Delta}{2} \hat{\sigma}_L^x \hat{\sigma}_{L+1}^x + \frac{J-\Delta}{2} \hat{\sigma}_L^y \hat{\sigma}_{L+1}^y = (-1)^{L+1}[\Delta(\hat{c}_L^\dagger \hat{c}_1^\dagger + \hat{c}_1 \hat{c}_L) + J(\hat{c}_L^\dagger \hat{c}_1 + \hat{c}_1^\dagger \hat{c}_L)]. \quad (11)
$$

And then according to the relation

$$
\hat{\sigma}_j^z = 2\hat{c}_j^\dagger \hat{c}_j - 1, \quad (12)
$$

substituting Eq. 8, Eq. 11 and Eq. 12 into Eq. 1, the quadratic expression with fermionic operators can be obtained as follows

$$
\begin{aligned}
\hat{H}_{ini} = &- \sum_j (J \hat{c}_j^\dagger \hat{c}_{j+1} + \Delta \hat{c}_j^\dagger \hat{c}_{j+1}^\dagger + \text{h.c.}) - \mu \sum_j \hat{c}_j^\dagger \hat{c}_j \\
&+ (-1)^L (\hat{c}_L^\dagger \hat{c}_1 + \hat{c}_1^\dagger \hat{c}_L + \Delta(\hat{c}_L^\dagger \hat{c}_1^\dagger + \hat{c}_1 \hat{c}_L)),
\end{aligned}
\quad (13)
$$

which is the Eq. (1) in the main text. Therefore, the APBC and periodic boundary conditions (PBC) should be used when the system size $L = 2m$ and $2m + 1$, respectively, where $m \in \mathbb{Z}$.

## SECTION II. DETAILS OF THE NUMERICAL IMPLEMENTATION

This section details our numerical approach and addresses potential sources of error to ensure result reliability. Simulating time evolution under the non-Hermitian Hamiltonian $\hat{H}_f$ requires careful numerical treatment. Specifically, the non-Hermitian nature can lead to exponentially growing numerical instabilities when calculating the time evolution operator $e^{-i\hat{H}_f t}$ directly over long times. To mitigate this, we employ a standard Trotter decomposition, dividing the total time $t$ into $N = t/\tau_0$ small steps: $e^{-i\hat{H}_f t} \approx (e^{-i\hat{H}_f \tau_0})^N$, where $\tau_0 = \hbar/J$ is the elementary time interval. While this decomposition reduces error accumulation per step and yields physically consistent results for smaller systems (e.g., $L = 32$) at standard precision (15 digits), it becomes insufficient for larger systems (e.g., $L = 64$) or stronger non-Hermiticity. Numerical investigations revealed that standard precision leads to significant inaccuracies in matrix operations (like multiplication and diagonalization) involving large non-Hermitian matrices, potentially yielding unphysical outcomes over time. We found that increasing the numerical precision progressively reduces these errors. By utilizing high-precision arithmetic (250 digits), our results remain reliable even for system sizes up to $L = 128$ and non-reciprocity $\gamma = 1.0$, ensuring the accurate capture of the dominant physical behavior. This underscores the necessity of high-precision calculations for robust simulations of large non-Hermitian quantum systems.

## SECTION III. AN ARGUMENT FOR THE TOTAL PARTICLE NUMBER AT LATE TIMES

Based on the preceding calculations, the total particle number of the system under OBC tends toward half-filling during long-term evolution, whereas it remains conserved under APBC. In this section, combining with some quantitative calculations, we give an argument to elucidate the evolution of total particle numbers under different boundaries.

### Section III-A. The case under open boundary condition

Under OBC, the steady-state value of the total particle number approaches half-filling as $\gamma$ increases. Below, we consider the limit $\gamma = 1$ to demonstrate that the total particle number tends towards half the total system size in the long-time evolution.

The evolution Hamiltonian is the Hatano-Nelson(HN) model

$$\hat{H}_{\text{evo}} = -\sum_j (t_L \hat{c}_j^\dagger \hat{c}_{j+1} + t_R \hat{c}_{j+1}^\dagger \hat{c}_j), \tag{14}$$

where $t_L = J + \gamma$, $t_R = J - \gamma$ represents the non-reciprocal hopping amplitudes. The state under the evolution operator is

$$|\psi(t)\rangle = e^{-i\hat{H}_{\text{evo}} t}|\psi(0)\rangle \tag{15}$$

The total particle number is

$$\begin{aligned}
\overline{N}(t) &= \frac{\langle \psi(t)|\hat{N}|\psi(t)\rangle}{\langle \psi(t)|\psi(t)\rangle} \\
&= \frac{\langle \psi(0)|e^{i\hat{H}_{\text{evo}}^\dagger t} \sum_l \hat{c}_l^\dagger \hat{c}_l e^{-i\hat{H}_{\text{evo}} t}|\psi(0)\rangle}{\langle \psi(0)|e^{i\hat{H}_{\text{evo}}^\dagger t} e^{-i\hat{H}_{\text{evo}} t}|\psi(0)\rangle},
\end{aligned} \tag{16}$$

When we set $\gamma = 1$, the Eq. 16 becomes

$$\overline{N}(t) = \frac{\langle \psi(0)|e^{2i\gamma \hat{H}_R t} \hat{N} e^{-2i\gamma \hat{H}_L t}|\psi(0)\rangle}{\langle \psi(0)|e^{2i\gamma \hat{H}_R t} e^{-2i\gamma \hat{H}_L t}|\psi(0)\rangle}, \tag{17}$$

where $\hat{H}_L = \sum_j \hat{c}_j^\dagger \hat{c}_{j+1}, \hat{H}_R = \sum_j \hat{c}_{j+1}^\dagger \hat{c}_j$. Using the Jordan-Wigner transformation, the initial state in the main text can be expressed in terms of spinless fermions as

$$|\psi(0)\rangle = \sum_n (-1)^{n(n-1)/2} a_n |n\rangle, \tag{18}$$

where $a_n = \sqrt{2}\cos^n\frac{\theta}{2}\sin^{(L-n)}\frac{\theta}{2}$, and $|n\rangle$ contains $C_L^n$ different configurations corresponding $n$ fermions occupation. The action of $\hat{H}_L$ on the state in $|n\rangle$ is to carry the fermion to the left nearest-neighbor until it encounters an occupied fermion or the boundary. Thus

$$\begin{aligned}e^{-2i\gamma\hat{H}_Lt}|\psi(0)\rangle &= \sum_n (-1)^{n(n-1)/2} a_n e^{-2i\gamma\hat{H}_Lt}|n\rangle \\ &= \sum_n (-1)^{n(n-1)/2} a_n (\sum_m \frac{(-2i\gamma t)^m}{m!}(\hat{H}_L)^m)|n\rangle \\ &= \sum_n (-1)^{n(n-1)/2} a_n \\ &\quad \times \underbrace{(\sum_{m_1}\frac{(-2i\gamma t)^{m_1}}{m_1!}(\hat{H}_L)^{m_1}|\text{\textcircled{1}}\rangle + \sum_{m_2}\frac{(-2i\gamma t)^{m_2}}{m_2!}(\hat{H}_L)^{m_2}|\text{\textcircled{2}}\rangle + \cdots)}_{C_L^n},\end{aligned} \quad (19)$$

where $|\text{\textcircled{n}}\rangle$ represents the $n$th configuration and $C_L^n$ is the number of configurations in state $|n\rangle$. For a fixed configuration $|\text{\textcircled{n}}\rangle$, there exists a minimum value $n_{min}$ such that $(\hat{H}_L)^{n_{min}+1}|\text{\textcircled{n}}\rangle = 0$, where $n_{min}$ corresponds to the maximum times these fermions can be carried to left side in a given configuration. We can find the fermions in the configuration $|\underbrace{00\cdots0}_{L/2}\underbrace{11\cdots1}_{L/2}\rangle$ can have the maximum times $n_{min} = (L/2)^2$ to be carried. Specifically,

$$\frac{(-2i\gamma t)^{(L/2)^2}}{(L/2)^2!}|\underbrace{00\cdots0}_{L/2}\underbrace{11\cdots1}_{L/2}\rangle \equiv C_{max}|\underbrace{00\cdots0}_{L/2}\underbrace{11\cdots1}_{L/2}\rangle \quad (20)$$

Therefore, both the numerator and denominator in Eq. 17 are polynomials about variable $t$. To obtain the steady value for long times, we consider the result at the limit $t \to \infty$. Then the result of $\lim_{t\to\infty}\bar{N}(t)$ depends only on the leading-order term in $t$, that is,

$$\lim_{t\to\infty}\overline{N}(t) = \frac{|C_{max}|^2 \frac{L}{2} a_{L/2}^2}{|C_{max}|^2 a_{L/2}^2} = \frac{L}{2}. \quad (21)$$

It should be noted that total size $L$ is taken as an even number so that particle number $n$ in Eq. 18 can only also take even numbers. The specific numerical calculations above take $L = 64 = 2*32$, but if $L$ is taken as $2\times$odd number, the efficient $a_{L/2} = 0$. For the case of $L = 2n$ ($n$ is an odd number) we need to consider the sub-leading term. We can find that the configurations $|\underbrace{00\cdots0}_{n+1}\underbrace{11\cdots1}_{n-1}\rangle$ and $|\underbrace{00\cdots0}_{n-1}\underbrace{11\cdots1}_{n+1}\rangle$ give the sub-leading term. Specifically,

$$e^{-i\hat{H}_Lt}|\psi(0)\rangle \propto \frac{(-2i\gamma t)^{(n-1)(n+1)}}{((n-1)(n+1))!}(a_{n-1}|\underbrace{00\cdots0}_{n+1}\underbrace{11\cdots1}_{n-1}\rangle + a_{n+1}|\underbrace{00\cdots0}_{n-1}\underbrace{11\cdots1}_{n+1}\rangle). \quad (22)$$

Therefore, we have

$$\begin{aligned}\lim_{t\to\infty}\overline{N}(t) &= \frac{(n-1)a_{n-1}^2 + (n+1)a_{n+1}^2}{a_{n-1}^2 + a_{n+1}^2} \\ &= n + \frac{2\cos\theta}{1+\cos^2\theta},\end{aligned} \quad (23)$$

that is,

$$\lim_{t\to\infty}\overline{N}(t)/L \to 1/2. \quad (24)$$

Through the verification of numerical calculation (taking $L = 62$), the steady values for the initial states with different tilted angles are consistent with Eq. 23.

**Section III-B. The case under (anti-)periodic boundary condition**

In this section, we focus on the total particle number in the long-time limit under PBC. Under PBC, the allowed momenta are given by $k \in \{m\pi/L | m = -L+1, \cdots, 0, \cdots, L\}$. (For APBC in the practical numerical calculation, the momentum is given by $k \in \{\pm(2m+1)\pi/L | m = 0, \cdots, L/2 - 1\}$, which doesn't affect the final results.)

Using the Fourier transformation

$$\hat{c}_k = \frac{1}{L} \sum_j \hat{c}_j e^{ikj}, \tag{25}$$

the HN model can be simplified to

$$\hat{H}_{\text{evo}} = \sum_k E_k \hat{c}_k^\dagger \hat{c}_k, \tag{26}$$

where the single-particle eigenvalues are

$$E_k = -2J\cos k - 2i\gamma \sin k \tag{27}$$

Since there is no interacting term in HN model, the many-body eigenstates $|\{n_k\}\rangle$ are Slater determinants formed by occupying a set $S(\{n_k\})$ of momenta $k$, where $n_k$ is the occupation number of momentum $k$ ($n_k = 0$ or $1$). The corresponding many-body eigenvalue is $E_{\{n_k\}} = \sum_{k \in S(\{n_k\})} E_k$. Therefore, the evolution of such an eigenstate is

$$\begin{aligned}
e^{-i\hat{H}_{\text{evo}}t}|\{n_k\}\rangle &= e^{-iE(\{n_k\})t}|\{n_k\}\rangle \\
&= e^{it\sum_{k \in S(\{n_k\})} 2J\cos k} e^{-t\sum_{k \in S(\{n_k\})} 2\gamma \sin k}|\{n_k\}\rangle.
\end{aligned} \tag{28}$$

For any initial state $|\psi(0)\rangle$, it can be expanded in the basis of these many-body states as

$$|\psi(0)\rangle = \sum_{\{n_k\}} a_{\{n_k\}} |\{n_k\}\rangle, \tag{29}$$

where $a_{\{n_k\}}$ are complex coefficients. Then we have

$$\begin{aligned}
|\psi(t)\rangle &= \sum_{\{n_k\}} a_{\{n_k\}} e^{-i\hat{H}_{\text{evo}}t}|\{n_k\}\rangle \\
&= \sum_{\{n_k\}} a_{\{n_k\}} e^{it\sum_{k \in S(\{n_k\})} 2J\cos k} e^{-t\sum_{k \in S(\{n_k\})} 2\gamma \sin k}|\{n_k\}\rangle
\end{aligned} \tag{30}$$

Substituting Eq. 30 into the numerator in Eq. 16, we have

$$\begin{aligned}
\langle \psi(t)|\hat{N}|\psi(t)\rangle &= \sum_{\{m_k\}} \sum_{\{n_k\}} a^*_{\{m_k\}} a_{\{n_k\}} e^{it(\sum_{k \in S(\{n_k\})} 2J\cos k - \sum_{k \in S(\{m_k\})} 2J\cos k)} \\
&\quad \times e^{-t(\sum_{k \in S(\{n_k\})} 2\gamma \sin k + \sum_{k \in S(\{m_k\})} 2\gamma \sin k)} \langle \{m_k\}|\hat{N}|\{n_k\}\rangle \\
&= \sum_{\{n_k\}} |a_{\{n_k\}}|^2 e^{-4\gamma t \sum_{k \in S(\{n_k\})} \sin k} N_{\{n_k\}}
\end{aligned} \tag{31}$$

where $N_{\{n_k\}}$ is the total number of particles in the state $|\{n_k\}\rangle$. Therefore, the total particle number is

$$\overline{N}(t) = \frac{\sum_{\{n_k\}} |a_{\{n_k\}}|^2 e^{-4\gamma t \sum_{k \in S(\{n_k\})} \sin k} N_{\{n_k\}}}{\sum_{\{n_k\}} |a_{\{n_k\}}|^2 e^{-4\gamma t \sum_{k \in S(\{n_k\})} \sin k}} \tag{32}$$

In the long-time limit, the component $\{n_k\}$ with all momenta $k \in (-\pi, 0)$ filled will dominate the summation. Here we denote it as $\{n_{k-}\}$. Assuming $a_{\{n_{k-}\}} \neq 0$, wa have

$$\begin{aligned}
\overline{N}(t) = \lim_{t \to \infty} \overline{N}(t) &= \frac{|a_{\{n_{k-}\}}|^2 N_{\{n_{k-}\}}}{|a_{\{n_{k-}\}}|^2} \\
&= \frac{L}{2}
\end{aligned} \tag{33}$$

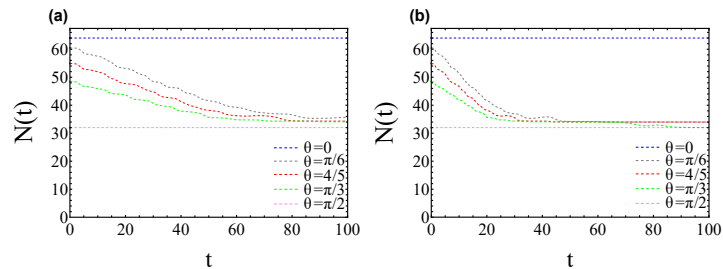

FIG. S1. The total particle number with the initial state generated by the initial Hamiltonian under OBC. (a) and (b) correspond to the results with $\gamma = 0.4$ and $1.0$, respectively. The legend represents different tilted angles and the total size $L = 64$.

Consequently, under PBC and after long-time evolution, the total particle number also tends towards half-filling, provided the initial state contains non-zero $\{n_{k-}\}$ components. For our initial state, which is a BCS state generated by Eq. (3) in the main text, $+k$ and $-k$ states exist in pairs in momentum space. Consequently, in the decay term of Eq. 32, the exponent $\sum_{k \in S(\{n_k\})} \sin k$ is equal to 0, leading to the conservation of the total particle number.

It is important to note that when imposing PBC on the initial Hamiltonian in Eq. (2) of the main text, the resulting initial state is a perfect BCS state. However, with OBC, the initial state is not a strict pairing state. The results presented in Fig. S10 are based on the initial state generated under PBC, hence the conservation of the total particle number. In contrast, for the results shown in Fig. S1, OBC are applied, and as can be seen, the total particle number indeed tends towards half-filling after long-time evolution.

## SECTION IV. QUASIPARTICLE PICTURE

This section gives more details for the quasiparticle picture to understand the entanglement dynamics.

### Section IV-A. Hermitian case

The quasiparticle picture provides a useful framework for understanding entanglement propagation in non-equilibrium processes. Here we first review the quasiparticle picture in the Hermitian case [1, 2].

In the Hermitian case, the quasiparticles at initial moment are assumed to exist in pairs with momentum $\pm k$ and uniformly distributed on every site. Subsequently, the pairs propagate to the left and right, respectively. A contribution to the increase in EE arises when a pair of quasiparticles, originating from the same spatial location, simultaneously reaches subsystems $A$ and $B$. Within this framework, the EE can be expressed as

$$S_{\text{vonEE}} \propto \int_{t \leqslant \frac{l}{2v(k)}} 2v(k) t s(k) \mathrm{d}k + \int_{t > \frac{l}{2v(k)}} l s(k) \mathrm{d}k, \tag{34}$$

where $l$ represents the subsystem size, $v(k)$ is the group velocity of the quasiparticle, and $s(k) = -n_k \ln(n_k) - (1 - n_k) \ln(1 - n_k)$ denotes the entanglement generation rate. During the time interval $0 \leqslant t \leqslant l/2v(k)$, quasiparticle pairs emitted from various locations gradually arrive at subsystems $A$ and $B$, leading to an increase in EE. Then for $t > l/2v(k)$, the majority of quasiparticle pairs reside within subsystems $A$ and $B$, causing the EE to approach saturation. The initial states characterized by different tilted angles result in distinct quasiparticle distributions, which are reflected in variations of $n_k$. Due to the isotropic propagation of the quasiparticles, the EE is generally independent of the subsystem's position.

### Section IV-B. Non-Hermitian case

Before delving into the non-Hermitian quasi-particle picture, we consider the effects of the skin mode on a single-particle wave packet. As shown in Ref. [3], for a Gaussian wave packet with initial position $x_0$ and negative initial momentum $k_0 < 0$, the skin mode has two primary consequences for its dynamics:

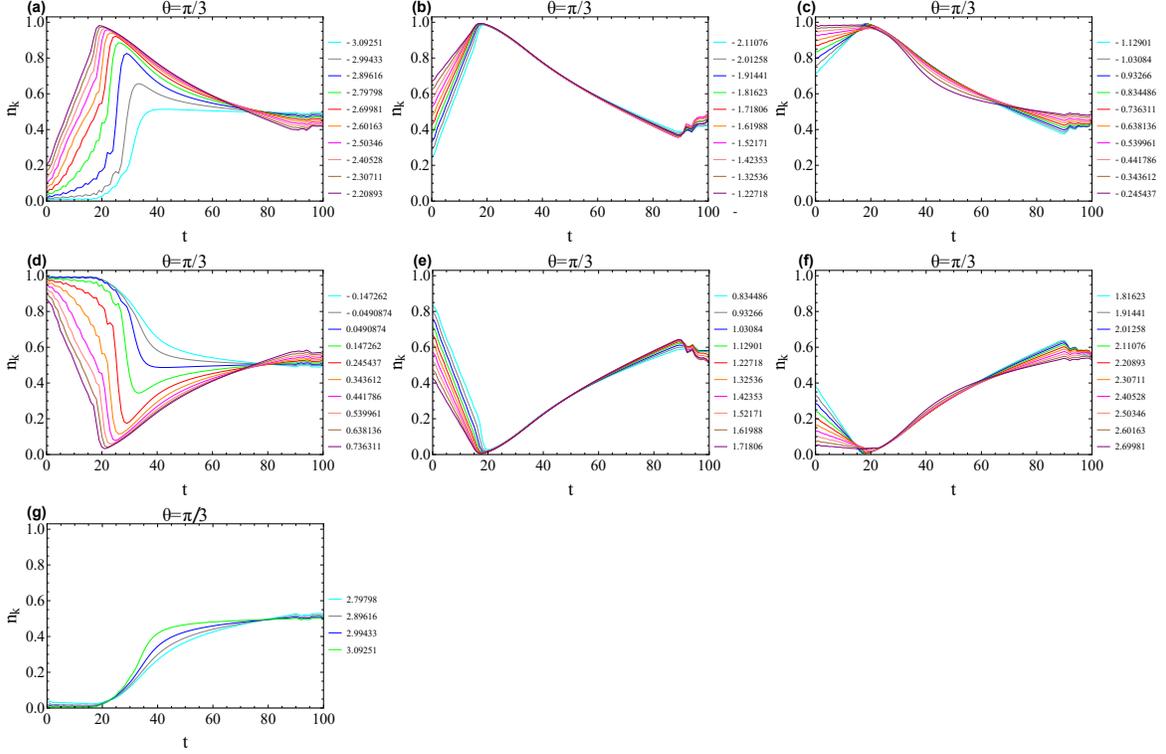

FIG. S2. The dynamics of particle density in the momentum space from the tilted states with $\theta = \pi/3$. The numbers in the legend in (a)-(g) represents the momentum $k = \pm(2j-1)\pi/L$, where the $j = 1, ..., L/2$. The total size $L = 64$ and non-reciprocal strength $\gamma = 0.8$.

1. The modulus of wave packet begins to decay from the initial time (we therefore normalize the wave function at each time); however, its momentum remains unchanged.

2. Upon reaching the boundary, wave packets with smaller initial momenta $k_0$ exhibit subsequent evolution that is localized near the boundary, exhibiting behavior akin to being stuck to the boundary, i.e. the sticky effect induced by the skin modes. In contrast, particles with larger initial momenta can follow the standard laws of reflection.

Building upon the aforementioned two characteristics, we now consider the quasiparticle picture in the non-Hermitian case. The first effect above, fundamentally a consequence of the non-unitary evolution, may persist even in the absence of the skin modes. A common assumption in previous studies under OBC has been that quasiparticles propagating towards the side where the skin mode is concentrated are amplified, while those propagating in the opposite direction are attenuated [4, 5]. This assumption corresponds to the first effect above but does not address the influence of the boundary on quasiparticle reflection behavior. Given the non-unitarity, quasi-particles emitted uniformly at each site exhibit anisotropic propagation after the initial time. During their propagation before reaching the boundary, their amplification or attenuation depends on their emission location. Thus, quasiparticle pairs with the same momentum $k$ will arrive at the boundary non-uniformly. Considering the second effect of the skin modes on a wave packet, the reflection behavior near the boundary may vary for each quasiparticle with momentum $k$. Thus, the question arises: How does the skin mode influence the reflection behavior of this ensemble of quasiparticles with momentum $k$?

On the other hand, the Hermitian picture dictates that EE increases only when quasiparticle pairs simultaneously reach subsystems $A$ and $B$, a manifestation of their correlation [1, 2]. We examine the evolution of the particle density $n_k$ in momentum space. Here we directly use the Fourier transformation in Eq. 25 to calculate the particle density in the momentum space $n_k = \langle \psi(t)|\hat{c}_k^\dagger \hat{c}_k|\psi(t)\rangle$, even though its physical interpretation is questionable. As shown in Fig. S2, quasiparticles with momentum $k < 0$ and $k > 0$ propagate to the left and right, respectively. During the time interval $0 < t < \tau_1$, quasiparticles located in the central region have not yet reached either boundary, resulting in $n_{k<0}$ increasing and $n_{k>0}$ decreasing monotonically with time, respectively. At the characteristic time $\tau_1$, the quasiparticle pairs with the maximum group velocity $v_{\text{gmax}}$ simultaneously reach both boundaries, leading to effective reflection of

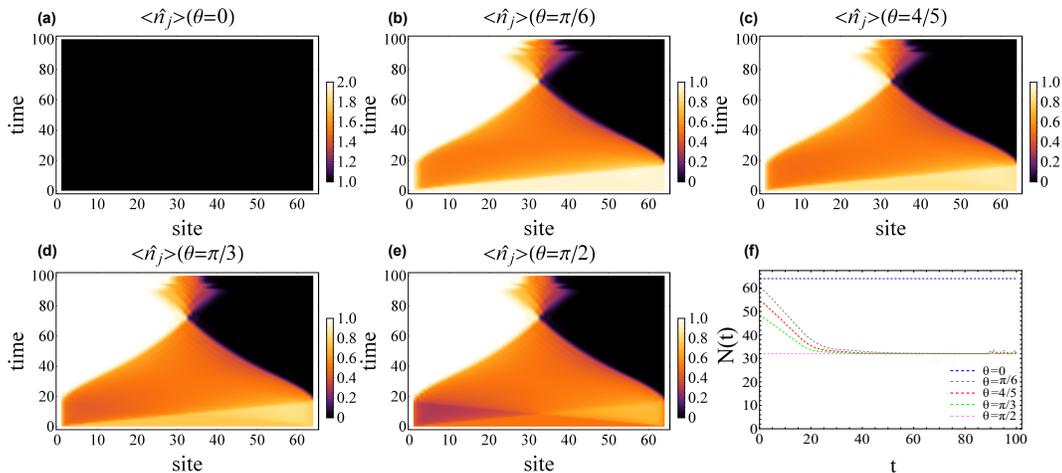

FIG. S3. The dynamics of on-site particle density in different tilted states with $\gamma = 0.8$ under OBC. The tilted angles $\theta$ are (a)0, (b)$\pi/6$, (c)4/5, (d)$\pi/3$, (e)$\pi/2$, respectively. (f) shows the evolution of total particle number corresponding to Fig.1 (a) in the main text. The total size $L = 64$.

quasiparticles at the boundaries. This reflection inverts the momentum sign, causing $n_{k<0}$ ($n_{k>0}$) to contribute to $n_{k>0}$ ($n_{k<0}$). Consequently, $n_{k>0}$, which was initially close to zero, starts to increase, while $n_{k<0}$ begins to decrease. Subsequently, the quasiparticles undergo multiple reflections between the two boundaries, eventually leading the system to a steady state with a 1/2 filling. Throughout the propagation process, quasiparticles originating from different locations exhibit different evolution behaviors, making the process non-uniform, thus invalidating Eq. 34. In section V-D, we directly utilize Eq. 34 to compute the EE, thereby illustrating the deviations from the Hermitian picture arising from the aforementioned non-uniform propagation.

## SECTION V. DYNAMICS UNDER OBC

This section elaborates on the dynamics of particle density and entanglement under OBC, providing further details to support the findings presented in the main text.

### Section V-A. The density dynamics with different parameters $\theta$ and $\gamma$

As illustrated in Fig. 1 of the main text, the density dynamics exhibit two key characteristics: the formation of a chiral wavefront in density, current, and inflow, and the eventual half-filling of the total particle number at steady state. Following the arrival of the chiral wavefront at the right boundary, the skin modes cause particle density to accumulate in the half of the system where the skin mode is concentrated, while the opposite half becomes nearly entirely depleted. In this section, we delve deeper into the influence of tilted angles $\theta$ and non-reciprocity $\gamma$ on these two universal features.

As illustrated in Fig. 1 in the main text, the on-site particle density exhibits an evolution that is nearly congruent with that of the current. Furthermore, the non-zero region of the inflow of particle is predominantly located at the boundary of the current. Therefore, we use the dynamics of the particle density as a representative example to investigate density dynamics. Fig. S3 illustrates the particle density dynamics for varying tilted states, while maintaining a fixed non-reciprocity strength of $\gamma = 0.8$. The ferromagnetic state ($\theta = 0$) in Fig. S3(a) is an eigenstate of $\hat{n}_j$, so the particle on every site is conserved. In Fig. S3(b)-(e), all tilted states with $\theta \neq 0$ have a wave front propagating from left to right with a velocity of approximately twice the maximum group velocity ($2v_{\text{gmax}}$). As $\theta$ increases, a secondary wavefront emerges, propagating from right to left. The trajectory of this wavefront becomes more pronounced with larger $\theta$. Notably, when $\theta$ reaches $\pi/2$, the trajectories of the two wavefronts become nearly symmetric with respect to the line $x = L/2 + 0.5$. Unlike the wavefront originating from the left, the particle density starts to increase at the site where the wavefront propagating from the right side arrives. Furthermore, comparing Fig. S3(e) and (f) reveals that the initial state with $\theta = \pi/2$ already exhibits half-filling, despite the on-site density not being conserved.

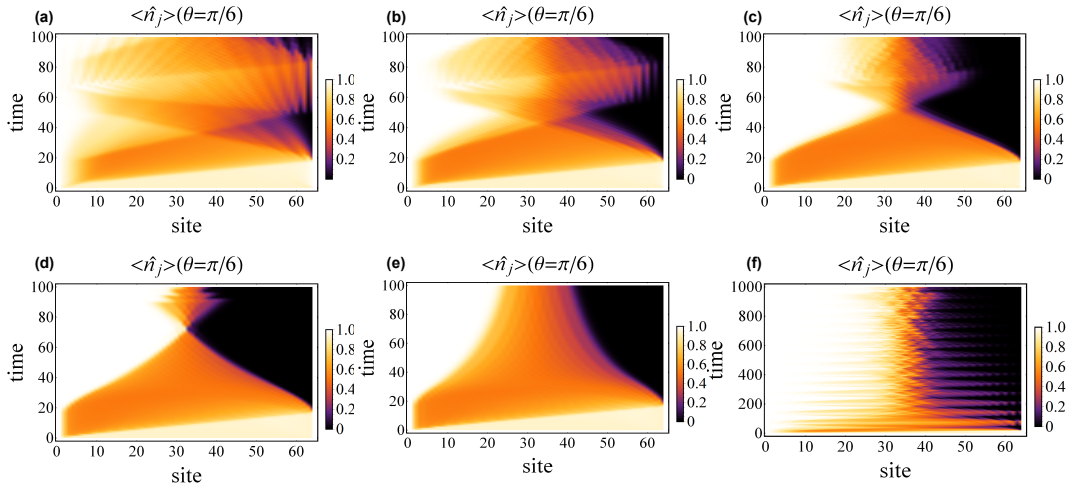

FIG. S4. The dynamics of on-site particle density in tilted state with $\theta = \pi/6$ under OBC. The non-reciprocity strength $\gamma$ are (a)0.2, (b)0.4, (c)0.6, (d)0.8, (e)1.0, respectively. (a)-(e) are the results of evolution in a time range of $0 \leq t \leq 100$, while (f) is the result in a range of $0 \leq t \leq 1000$ with $\gamma = 0.2$. The total size $L = 64$.

The main text highlights that at steady state, the system achieves half-filling, with particles accumulating on the side where the skin mode is localized, irrespective of the tilted angle. Here, we examine the influence of varying non-reciprocity strengths $\gamma$ on the particle density dynamics, using a tilted state with $\theta = \pi/6$ as a representative example. Regardless of the value of $\gamma$, universal chiral wavefront emerges in the initial dynamics, propagating from left to right with a velocity of $2v_{\text{gmax}}$. As mentioned in the main text, two characteristic times, $\tau_1$ and $\tau_2$, play a crucial role. $\tau_1 \approx L/(2v_{\text{gmax}})$ corresponds to the time when the chiral wavefront reaches the right boundary. After $\tau_1$, the particle density on both sides begins to incarese and decreases from the left and right side to the middle part, respectively, until the two fronts meet at time $\tau_2$. The characteristic time $\tau_1$ remains relatively constant, while $\tau_2$ monotonically increases with $\gamma$. For smaller values of $\gamma$, such as 0.2 shown in Fig. S4(a), the particle density exhibits larger oscillation amplitude after $\tau_2$. As demonstrated in Fig. S4(f), which displays the dynamics with $\gamma = 0.2$ over a longer time range (from 0 to 1000), the oscillation amplitude gradually decreases after prolonged evolution. As $\gamma$ increases, the oscillations in the particle density after $\tau_2$ gradually diminish, leading to a faster convergence to the steady state.

### Section V-B. The particle density on each site

To further visualize the chiral wavefront and the half-filling steady state, Fig. S5 presents the particle density at each site as a function of time. Initially, as time progresses, the particle density at each site successively decreases towards approximately 0.5, remaining relatively stable until the characteristic time $\tau_1$. After $\tau_1$, the particle density of sites in the left half rises successively to approximately 1.0, while the sites in the right half fall successively to zero. Around time $\tau_2$, the particle density near the middle position shows a significant oscillation.

### Section V-C. The density dynamics of tilted antiferromagnetic states

In this section, we present the particle density dynamics using an initial antiferromagnetic state with a fixed number of particles. While its long-time behavior is consistent with that of tilted ferromagnetic states, a chiral wavefront is absent at short times. Here, we further calculate the particle density dynamics of the tilted antiferromagnetic state with $\theta = \pi/6$ as an example, as shown in Fig. S6. The chiral wavefront is absent in the particle density dynamics at short times, which is independent of the value of $\gamma$. Similar to the results in Fig. S4, the oscillation amplitude in the middle region gradually decreases as $\gamma$ increases. Furthermore, the total particle number of the system is conserved during the evolution, regardless of the tilted angle.

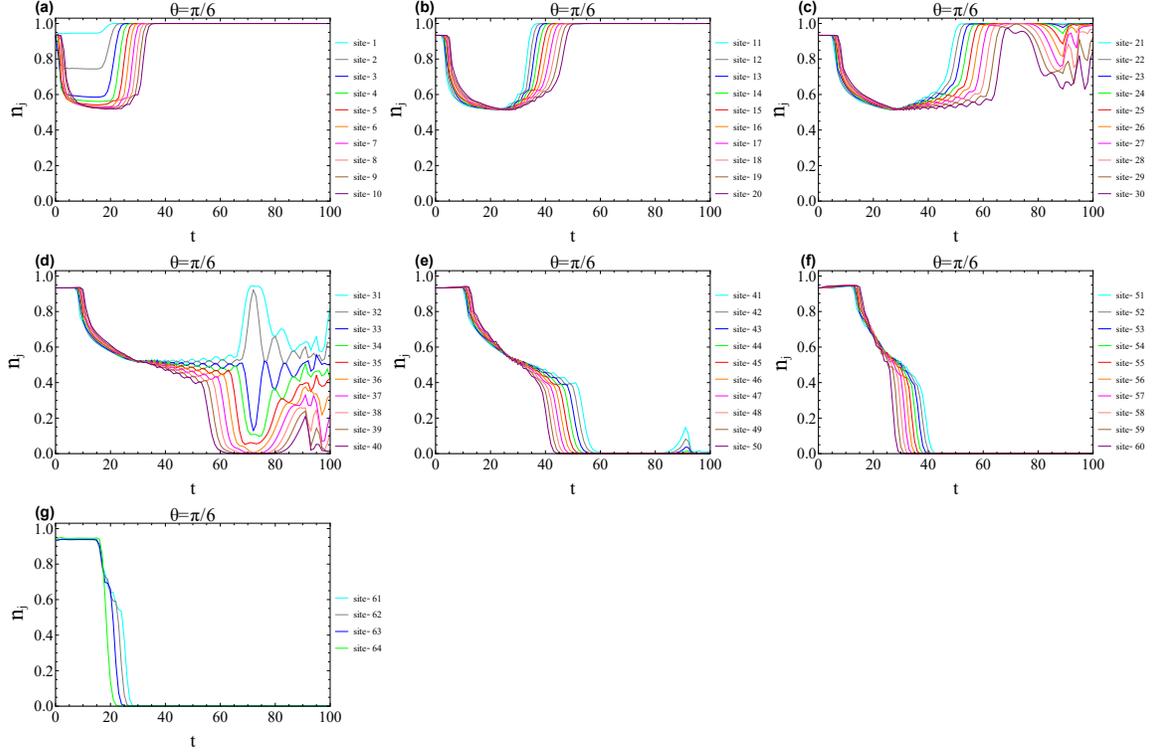

FIG. S5. The dynamics of on-site particle density in the tilted state with $\theta = \pi/6$ under OBC. The numbers in the legend in (a)-(g) from 1 to 64 label the positions of lattice site, where the total size $L = 64$. The non-reciprocal strength $\gamma = 0.8$.

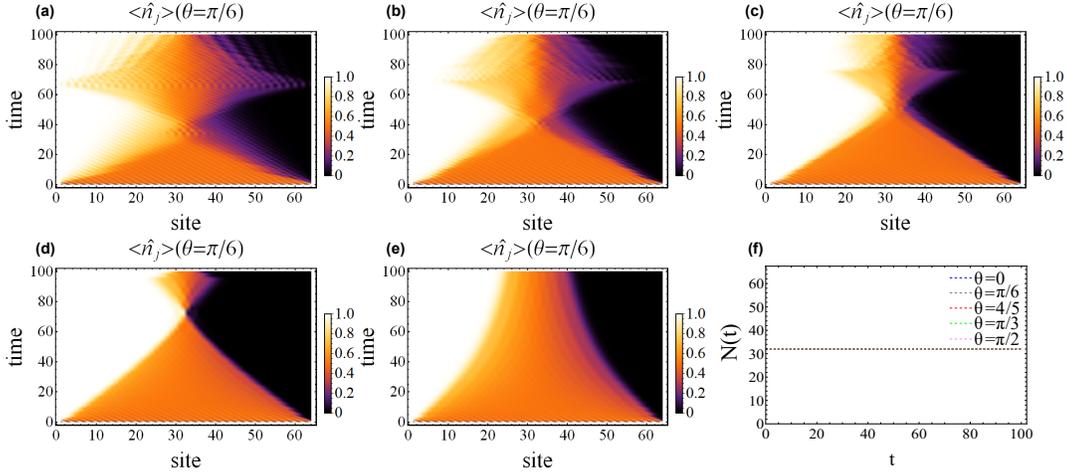

FIG. S6. The dynamics of on-site particle density in tilted antiferromagnetic state with $\theta = \pi/6$ under OBC. The non-reciprocity strength $\gamma$ are (a)0.2, (b)0.4, (c)0.6, (d)0.8, (e)1.0, respectively. (f) shows the evolution of total particle number, where the legends represent different tilted angles. The total size $L = 64$.

### Section V-D. The dynamics of entanglement entropy

This section provides further discussion on the entanglement dynamics presented in the main text. Here we show the dependence of entanglement dynamics on the position of the traced subsystem $B$. In the Hermitian case, as illustrated in Fig. S7(a), the EE for different tilted states exhibits similar evolutionary behaviors. Based on the quasiparticle picture in section IV-A, during the time interval $0 \leqslant t \leqslant l/2v(k)$, quasiparticle pairs emitted from various locations simultaneously reach subsystems $A$ and $B$, resulting in an increase in EE. Subsequently, for $t > l/2v(k)$, the majority

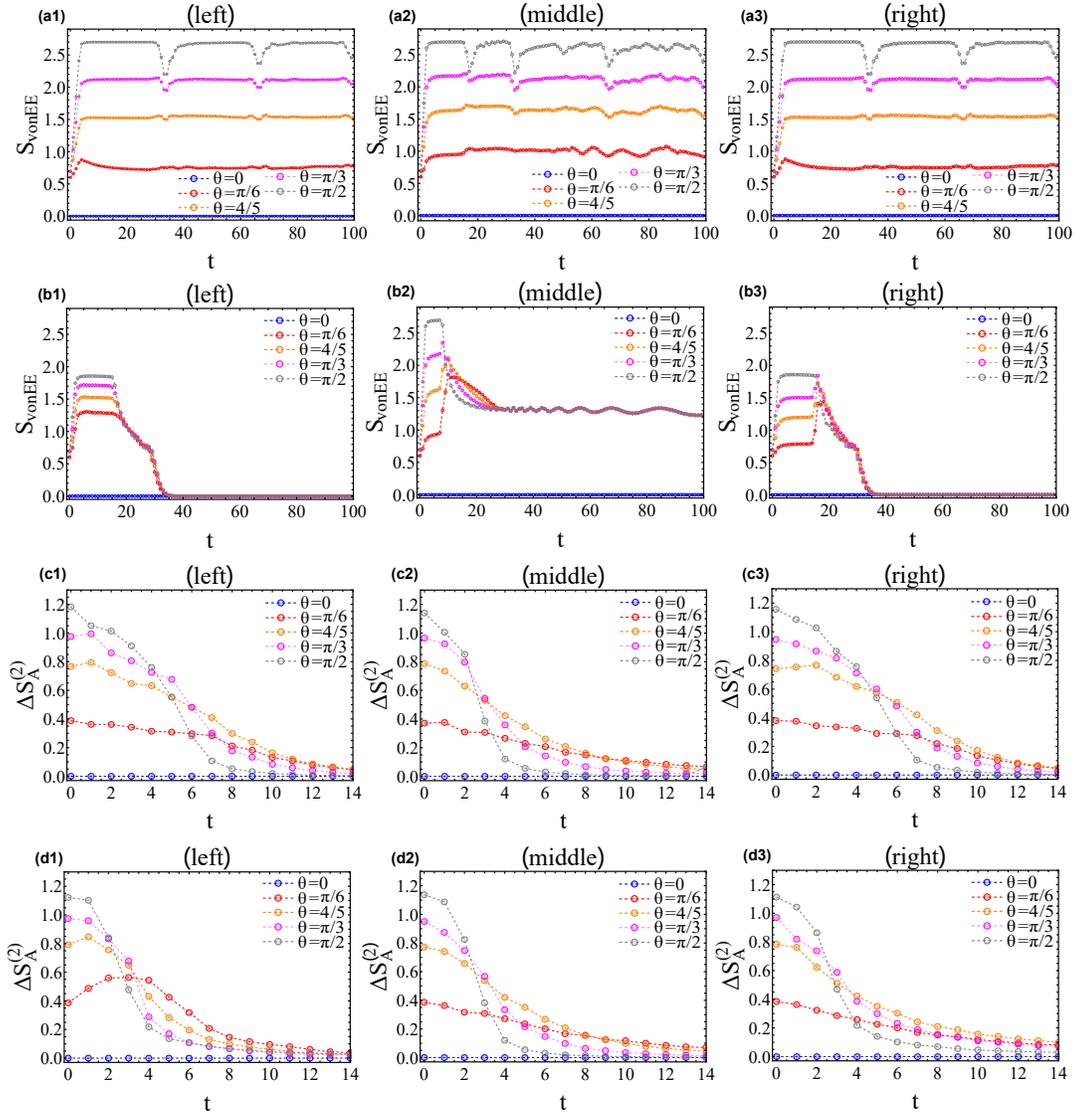

FIG. S7. (a)-(b)/(c)-(d) show the dynamics of EE/EA under OBC, where $\gamma = 0.0(0.8)$ in (a) and (c)((b) and (d)). The different colors represent different tilted angles. The label (left), (middle) and (right) correspond to the position of traced subsystem $B$. The total size $L = 64$, and the traced subsystem $B$ size in (a)-(b) and (c)-(d) are $l = 6$ and $12$, respectively.

of these pairs remain simultaneously present in both subsystems $A$ and $B$, causing the EE to approach saturation. The distinct quasiparticle structure arising from different tilted states, manifested in variations in the particle density $n_k$, impacts both the rate of EE growth and its eventual saturation value. Specifically, the numerical results in Fig. S7(a) demonstrate a monotonic increase in the initial EE growth rate with increasing tilted angle, while the saturation time remains approximately consistent across different tilted angles. Given that quasiparticles are uniformly produced at each site and propagate isotropically throughout the system, the EE evolution remains largely invariant with respect to the location of the subsystems.

In contrast, the results for the non-Hermitian case, shown in Fig. S7(b), the growth rate of EE in the initial stage, the maximum value and the steady state over a long time all depend on the location of the subsystem. Within the time interval $0 < t < \tau_1$, quasiparticles propagate anisotropically in two opposing directions. Due to the lack of effective quasiparticle reflection at the boundaries, which are dominated by the skin modes, the EE initially increases and then saturates, a behavior similar to that observed in the Hermitian case. As shown in Fig. S7(b), when the traced subsystem $B$ is located on either the left or right side, the EE exhibits similar behavior; however, when located on the left, the EE initially increases at a faster rate. Subsequently, around the characteristic time $t \approx \tau_1$, the skin mode begins to drive particle accumulation towards the left while simultaneously causing particle decay on the right,

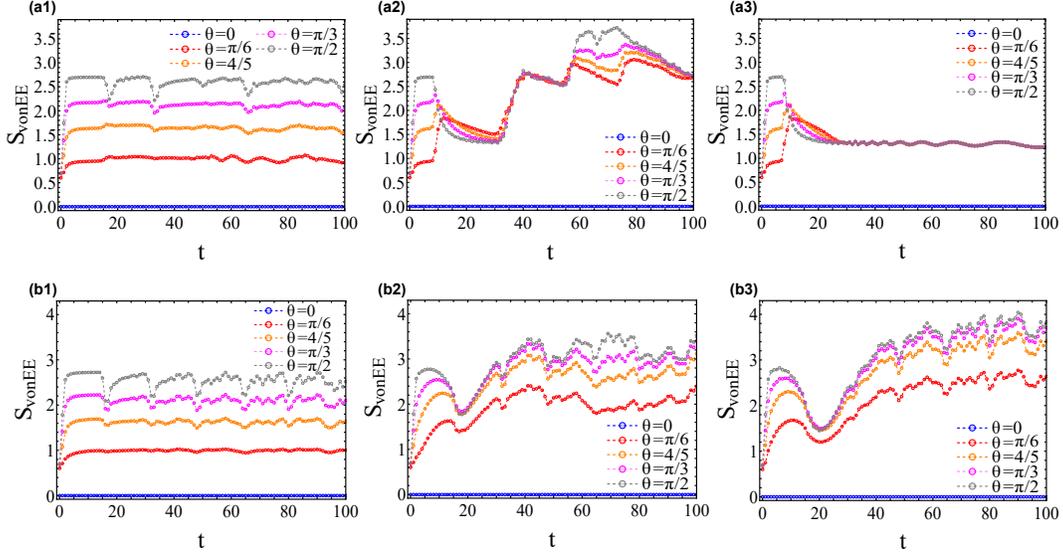

FIG. S8. (a) and (b) are the dynamics of EE under OBC, obtained from numerical calculation and application of Hermitian quasiparticle picture, respectively. The first, second, and third columns correspond to $\gamma = 0.0$, $0.2$ and $1.0$, respectively. Different colors in the legend represent different tilted angles. The traced subsystem $B$ with length $l = 6$ is located at the middle part of the system. The total system size is $L = 64$.

leading to a product-state-like behavior on both sides. Therefore, the number of quasiparticles contributing to EE gradually decreases, leading to a reduction in EE, potentially diminishing to zero in the steady state. However, when the subsystem $B$ is located in the middle position, the EE of each tilted state increases at a rate comparable to the Hermitian case, but its maximum begins to decrease earlier than $\tau_1$. The existence of small-amplitude oscillations in the particle density within the central region prevents the EE from reaching zero, leading to a small, non-zero steady-state value.

Referring back to the discussion in section IV-B, the quasiparticles emitted from different locations exhibit distinct behaviors during their propagation, rendering the system inhomogeneous. Consequently, Eq. 34 is no longer applicable. Therefore, we directly use the Fourier transformation in Eq. 25 and the Eq. 34 in Hermitian picture to calculate the EE, aiming to check the deviation from the non-Hermitian results. Fig. S8(a) and (b) show EE for different $\gamma$'s, corresponding to the numerical results and those obtained using the Hermitian quasiparticle picture, respectively. The two sets of results agree well in Fig. S8(a1) and (b1) when $\gamma = 0.0$. However, as $\gamma$ increases, the short-time dynamics remain qualitatively consistent in Fig. S8(a2)/(a3) and (b2)/(b3), while the long-time behavior diverges significantly. Therefore, a more quantitative non-Hermitian description of the quasiparticle picture is necessary, which is left to study later.

### Section V-E. The dynamics of entanglement asymmetry

This section provides further discussion on the dynamics of EA presented in the main text. Similar to EE, EA also exhibits a dependence on the location of the subsystem. In the Hermitian case depicted in Fig. S7(c), the EA during the restoration of U(1) symmetry exhibits the Mpemba effect. The results for EA are largely consistent across different subsystem locations. However, for the non-Hermitian case shown in Fig. S7(d), the EA is also dependent on the location of the subsystem. Specifically, when subsystem $B$ is located on the left side, the EA for partially tilted states in the initial stages exhibits a non-monotonic behavior, indicating a continued breaking of the U(1) symmetry, but this phenomenon does not occur when the subsystem $B$ is located in the middle or on the right side. Nevertheless, the U(1) symmetry is ultimately fully restored for all tilted states, as shown in Fig. S7(d1)-(d3). A clear Mpemba effect is still observed during the reduction of EA, where initial states with larger U(1) symmetry breaking recover the symmetry in a shorter time. Compared with Fig. S7(d2)-(d3), the non-monotonic behavior of EA in (d1) makes the Mpemba effect more pronounced, and the total time required for all tilted states to recover U(1) symmetry is slightly shorter.

As mentioned in the main text and the content above, the EA dynamics of the partially tilted states exhibit

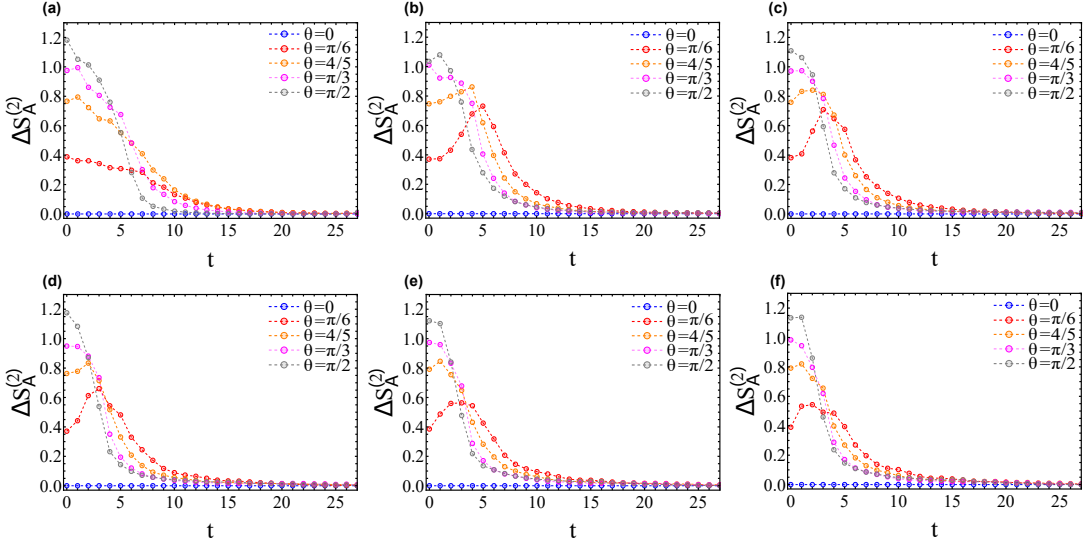

FIG. S9. The dynamics of EA under OBC with $\gamma = 0.0$ (a), 0.2 (b), 0.4(c), 0.6(d), 0.8(e) and 1.0(f). Different colors in the legend represents different tilted angles. The traced subsystem $B$ with length $l = 12$ is located at the left end of the system. The total system size is $L = 64$.

non-monotonic behavior at short times, characterized by an initial breaking of U(1) symmetry to a maximum value, followed by a recovery. Next, we explore the effect of non-reciprocity on this non-monotonic behavior. In the Hermitian case shown in Fig. S9(a), the evolution of all tilted states is generally decreasing. In Fig. S9(b)-(f) with different $\gamma$'s, the initial states with $\theta = \pi/6$ and $4/5$ initially increase and then decrease, indicating a breaking of U(1) symmetry to a maximum value. As $\gamma$ increases from 0.2 in Fig. S9(b), the magnitudes of two maxima gradually decrease. For $\gamma = 1.0$ in Fig. S9(f), the non-monotonic behavior corresponding to $\theta = 4/5$ diminishes significantly, while the maximum value for $\theta = \pi/6$ remains noticeable.

Therefore, non-reciprocity might enhance the superconductivity. A critical value for $\gamma$ likely exists within the range of 0 to 0.2, potentially leading to a maximum enhancement of superconductivity. In addition, compared to the Hermitian case in Fig. S9(a), the Mpemba effect is more pronounced in Fig. S9(b) with $\gamma = 0.2$. However, as $\gamma$ increases, the different tilted states gradually collapse together during the restoration of U(1) symmetry. Similar to the potential enhancement of superconductivity, non-reciprocity may also enhance the Mpamba effect, suggesting the existence of another critical value for $\gamma$ between 0 and 0.2, at which the Mpemba effect is most pronounced.

## SECTION VI. THE DYNAMICS WITH APBC

Section V provide a more detailed supplement to the dynamics under OBC discussed in the main text. In this section, we focus on the system's dynamics under APBC. APBC is employed for the Hamiltonian in Eq. 13 because the system size is even ($L = 64$), resulting in a factor of $(-1)^{L+1}$ when the Jordan-Wigner transformation is applied to the Hamiltonian with PBC in Eq. 1.

Regarding particle dynamics, Fig. S10(a) demonstrates that the total particle number is conserved for different tilted states, a point rigorously proven in section III. Furthermore, the particle density at each site is also conserved, as illustrated in Fig. S10(b) with $\theta = \pi/6$ as an example. Examination of the current at each site reveals a null result, indicating the absence of net current in the system, as shown in Fig. S10(c). Therefore, under APBC, the continuity equation of particle density holds, and there is no particle inflow. In the Hermitian case, the particle density at each site is also conserved, with no net current or particle inflow. Consequently, the non-Hermitian density dynamics under APBC are almost identical to those in Hermitian case.

Next, we examine the entanglement dynamics under APBC. Both the EE and EA evolution closely resemble the Hermitian case. Figs. S11(a1) and (a2) show nearly identical EE dynamics, while Figs. S11(b1) and (b2) demonstrate very similar EA behavior. This strong similarity to the Hermitian case means that characteristic features, such as the Mpemba effect observed in the EA decay, naturally persist during non-reciprocal evolution under APBC.

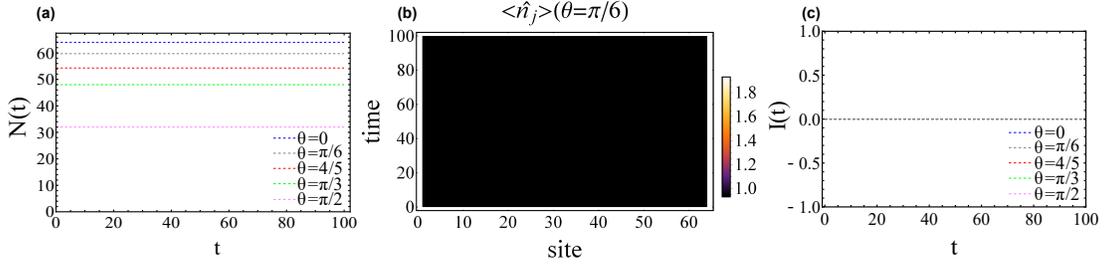

FIG. S10. Density dynamics under APBC. (a)The dynamics of total particle number with different tilted states represented by legends. (b)The dynamics of on-site particle density in titled state with $\theta = \pi/6$. (c)The dynamics of total current with different tilted states represented by legends. The total size $L = 64$.

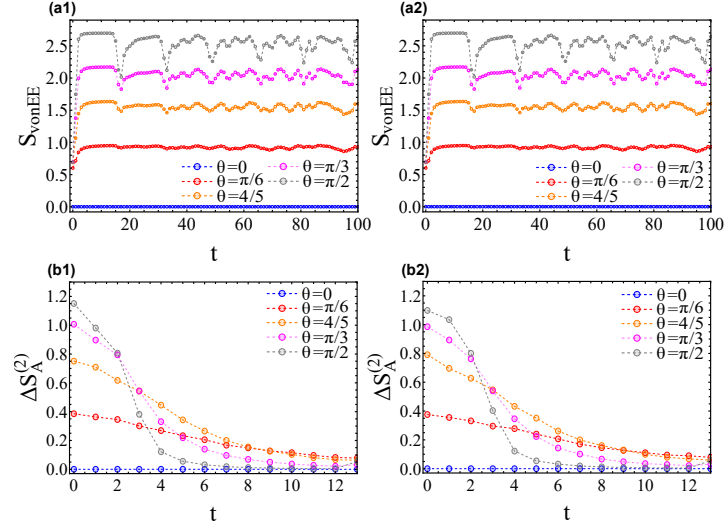

FIG. S11. (a) and (b) show the dynamics of EE and EA under APBC, respectively. The first and second column represent the Hermitian case and non-Hermitian case with $\gamma=0.8$, respectively. The legend represents different tilted angles and the total size $L = 64$. The traced subsystem $B$, located at the middle part of the system, has a size $l$ set to 6 and 12 in (a) and (b), respectively.

* shixinzhang@iphy.ac.cn
† zixiangli@iphy.ac.cn



\end{document}